%% file: ExtendedsubmissionWiOpt15.tex
\documentclass[10pt, conference, letterpaper]{IEEEtran}

\usepackage{amssymb}
\usepackage{amsthm}
\usepackage{caption}
\usepackage{cleveref}
\usepackage{colortbl}
\usepackage{comment}
\usepackage{epsfig}
\usepackage{graphicx}
\usepackage{listings}
\usepackage{multirow}
\usepackage{paralist}
\usepackage{rotating}
\usepackage{subcaption}
\usepackage{tabularx}
\usepackage[hyphens]{url}
\usepackage[usenames,dvipsnames]{xcolor}
\usepackage{xspace}

\usepackage{enumitem}

\usepackage{ amsfonts}
\usepackage{ amsthm}
\usepackage{epsfig}
\usepackage{cite}
\usepackage{algorithmic,algorithm}

\usepackage{setspace} 

\usepackage[english]{babel}

\usepackage[cmex10]{amsmath}
\usepackage{amsfonts,amssymb}

\usepackage{bm}
\usepackage{array}

\usepackage{color}

\begin{document}
\title{ Optimizing User Association and   Activation Fractions in Heterogeneous Wireless Networks}
\author{\IEEEauthorblockN{Vaibhav Singh, Narayan Prasad,
Mustafa Arslan, Sampath Rangarajan}
e-mail:   vaibhavs@umd.edu, \{prasad, marslan, sampath\}@nec-labs.com
}
\maketitle
\crefname{figure}{Fig.}{Figures}
\crefname{equation}{Equation}{Equations}
\crefname{table}{Table}{Tables}
\crefname{section}{Section}{Sections}
\crefname{subfigure}{Figure}{Figure}
\crefname{algocf}{Algorithm}{Algorithms}
\crefname{algorithm}{Algorithm}{Algorithms}

\input{Associationv2wbnE}

\input{GPwnE}
\input{jointlbasscv2wn}
\input{Evaluationv2wn}


%
%
 \appendix
\section{Appendix: Definitions}\label{app:defns}
We capture some basic definitions that are used in this paper.

\begin{definition}
Given a ground set $\Omega$, we define its power set  (i.e., the set containing all the subsets of  $\Omega$) as $2^{\Omega}$. Then,
a real-valued function defined on the subsets of $\Omega$, $h:2^{\Omega}\to\Reals$  is normalized if $h(\phi)=0$, where $\phi$ denotes the empty set. It is called a
{\em submodular} set function if and only if
\begin{eqnarray*}
h(\Bc\cup a)-h(\Bc)\leq h(\Ac\cup a)-h(\Ac),\\\;\forall\;\Ac\subseteq\Bc\subseteq\Omega\;\&\;a\in\Omega\setminus\Bc
\end{eqnarray*}
and a
{\em supermodular} set function if and only if
\begin{eqnarray*}
h(\Bc\cup a)-h(\Bc)\geq h(\Ac\cup a)-h(\Ac),\\\;\forall\;\Ac\subseteq\Bc\subseteq\Omega\;\&\;a\in\Omega\setminus\Bc.
\end{eqnarray*}
A non-negative valued set function $h:2^{\Omega}\to\Reals_+$ is a   non-decreasing set function if and only if it satisfies, $0\leq h(\Ac)\leq h(\Bc),\;\forall\;\Ac\subseteq\Bc\subseteq\Omega$.
\end{definition}
\begin{definition}
 $(\Omega,\Iul)$ is said to be
a {\em partition matroid} when   there exists a partition $\Omega=\cup_{i=1}^J\Omega_i$, where $\Omega_i\cap\Omega_j=\phi,\;\forall\;i\neq j$, along with integers $n_i\geq 1\;\forall\;i$ such that
\begin{eqnarray}
 \Bc\subseteq\Omega: |\Bc\cap\Omega_i|\leq n_i\;\forall\;i \Leftrightarrow \Bc\in\Iul.
  \end{eqnarray}
\end{definition}

\textbf{Proof of (8)}

 We will show in brief that for each TP $b\in\Bc$
 \begin{eqnarray}
\begin{split}
  &\mbox{max}_{\gamma_{k,b}\in[0,1]\;\forall\;k\atop\sum_{k\in\Uc}\gamma_{k,b}= 1}\left\{\sum_{k\in\Uc}x_{k,b}\left( w_k u(\gamma_{k,b}R_{k,b}(\rhob) )\right)\right\}
 = \\&- \left(\sum_{k\in\Uc}x_{k,b}\left(w_k \frac{(R_{k,b}(\rhob) )^{1-\alpha}}{\alpha-1}\right)^{1/\alpha}\right)^{\alpha}
\end{split}
\end{eqnarray}
The lagrangian for the convex optimization problem stated above  is given by \\
\begin{eqnarray}\label{eq:lagrangian}
\begin{split}
&\sum_{k\in\Uc}\frac{-x_{k,b} w_k {(\gamma_{k,b}R_{k,b}(\rhob))}^{1-\alpha}}{\alpha-1}+\sum_{k\in\Uc}{\lambda_{k}(\gamma_{k,b})}\\&+\mu(1-\sum_{k\in\Uc}{\gamma_{k,b}})
\end{split}
\end{eqnarray}
Using the first order derivative conditions and complementary slackness conditions, it is seen that
 the objective attains maximum value when for each user $k:x_{k,b}=1$, $\lambda_k=0$ so that $\gamma_{k,b}>0$, and  the following conditions are satisfied.\\
\begin{eqnarray}\label{eq:conditions}
\begin{split}
&  w_k {(\gamma_{k,b})}^{-\alpha}{R_{k,b}(\rhob)}^{1-\alpha}=\mu,\;\forall\;k:x_{k,b}=1;\\
& \sum_{k\in\Uc:x_{k,b}=1}\gamma_{k,b}= 1
\end{split}
\end{eqnarray}
Solving for optimal $\gamma_{k,b}$ from (\ref{eq:conditions}) and putting its value back in the objective, we obtain the RHS of (\ref{eq:tporiginal}).\\


\textbf{Proposition 1}:\\
{\em{Hardness of User Association}:}
The hardness of the user association sub-problem for a fixed $\rhob$ can be shown via a reduction from the {\em partition problem}.
To show this, we consider the case $\alpha>1$ and suppose that there is an optimal polynomial time user association algorithm. Further, we restrict ourselves to input instances in which the rates that all users can obtain from two distinct TPs $b1,b2\in\Bc$ are identical to one, whereas
  the rate that each user can obtain from any other TP is zero. Thus, we assume that
 $R_{k,b}(\rhob)=1,\;\forall\;k\in\Uc\;\&\;b\in\{b1,b2\}$ while $R_{k,b}(\rhob)=0,\;\forall\;k\in\Uc\;\&\;b\in\Bc\setminus\{b1,b2\}$. We allow the user weights to be any input set of positive scalars that sum to 1.
 Then, the problem in (\ref{eq:bg1}) simplifies to
 \begin{eqnarray}\label{eq:bgred1}
\begin{split}
&\mbox{min}_{x_{k,b}\in\{0,1\}\;\forall\;k\in\Uc,b\in\{b1,b2\}\atop\sum_{b\in\{b1,b2\}}x_{k,b}=1\;\forall\;k }\left\{\sum_{b\in\{b1,b2\}} \left(\sum_{k\in\Uc}x_{k,b}w_k^{1/\alpha}\right)^{\alpha}\right\}.
\end{split}
\end{eqnarray}
 Then, defining $\hat{z}=\arg\min_{z\in[0,1]}\{z^{\alpha}+(1-z)^{\alpha}\}$, it is readily verified that $\hat{z}$ is unique and equal to $1/2$, with $\hat{z}^{\alpha}+(1-\hat{z})^{\alpha}=2^{1-\alpha}$.
 Letting $W=\sum_{k\in\Uc}w_k^{1/\alpha}$, this implies that the objective value in (\ref{eq:bgred1}) returned by the optimal polynomial time user association algorithm will be equal to $W^{\alpha}2^{1-\alpha}$ if and only if there exists a partition of the
  set of user weights (each raised to power $1/\alpha$) into two parts that have an identical sum. This in turn implies that the algorithm at hand is an optimal polynomial time algorithm for the NP-complete  partition  problem. Indeed, suppose $\{y_1,\cdots,y_K\}:\;y_k> 0,\;\forall k$ is any input set to the latter problem where we need to determine if there exists a partition of that set into two parts of identical sum.  Setting
   $w_k=\frac{y_k^{\alpha}}{\sum_{i=1}^Ky_i^{\alpha}},\;\forall\;k=1,\cdots,K$, we obtain a valid input set of weights for (\ref{eq:bgred1}). Then, from the output of the supposed optimal algorithm at hand, we can immediately determine
    if there is such a partition for the set $\{\frac{y_k}{(\sum_{i=1}^Ky_i^{\alpha})^{1/\alpha}}\}_{k=1}^K$ and thus the set $\{y_k\}_{k=1}^K$, which yields the desired contradiction.
  The same reduction can be established for $\alpha=1$ as well as $\alpha\in (0,1)$.

To prove the remaining parts of this proposition, we note  that $x^{\alpha}$ for all non-negative $x$ is concave in $x$ when $\alpha\in (0,1)$ and convex in $x$ when $\alpha>1$.  Then, we note the fact that composition of a non-negative modular set function with a  concave (convex) function yields a submodular (supermodular) set function. Further, submodularity as well as supermodularity is preserved under set restriction and the sum of submodular (supermodular) functions is submodular (supermodular). Using these facts,  we obtain the desired results. Similarly, for $\alpha=1$ we note that
 $-x\ln(x)$ is concave in $x$ for all non-negative $x$. This fact along with the aforementioned arguments and the fact that the sum of a submodular set function and a modular set function is submodular, establishes the proof in this case. Finally, since we allow for arbitrarily small (albeit positive) $R_{k,b}(\rhob)$ for any tuple $(k,b)$ the set function $g(.,1)$ need not be non-decreasing nor non-negative.

Before we consider Proposition 2, we state and prove a   lemma that will invoked later. The  bounds given in this lemma are  applicable to arbitrary  submodular or  supermodular set functions.

\textbf{Lemma 1}:\\
For any given $\alpha$,  the greedy stage yields an output $\hat{\Gulc}$ such that
\begin{eqnarray}\label{eq:gineq0}
\begin{split}
&g(\hat{\Gulc},\alpha)\geq g(\Gulc^{\rm opt}\cup\hat{\Gulc},\alpha)-g(\hat{\Gulc}\setminus\Gulc^{\rm opt},\alpha),\;\;\forall\;\alpha\in (0,1]\\
&g(\hat{\Gulc},\alpha)\leq g(\Gulc^{\rm opt}\cup\hat{\Gulc},\alpha)-g(\hat{\Gulc}\setminus\Gulc^{\rm opt},\alpha),\;\;\forall\;\alpha>1.
\end{split}
 \end{eqnarray}
  \proof
  We prove the first relation in (\ref{eq:gineq0}).
For notational convenience let us denote a tuple as $\eul=(k,b)$. We expand $\hat{\Gulc}$ as $\hat{\Gulc}=\{\hat{\eul}_1,\hat{\eul}_2,\cdots,\hat{\eul}_K\}$ where $\hat{\eul}_i$ denotes the tuple added at the $i^{th}$ greedy step and
let $\delta_i,\;i=1,\cdots,K$ denote the associated change in system utility. 
 Further, we define
 the sets $\hat{\Gulc}_i=\{\hat{\eul}_1,\hat{\eul}_2,\cdots,\hat{\eul}_i\},\;\forall\;\;i=1,\cdots,K$ with $\hat{\Gulc}_0=\phi$.
Then, note that both $\Gulc^{\rm opt},\hat{\Gulc}\in\Iulk$ and are   maximal members in $\Iulk$, i.e., $|\Gulc^{\rm opt}|=|\Gulc|=K$. Further, using the definitions given above, we see that $\Iulk$ is a  partition matroid. Invoking a result on maximal members in a matroid (cf. \cite{lee:nmsub}), we can deduce that without loss of generality, we can expand $\Gulc^{\rm opt}=\{\eul^{\rm opt}_1,\eul^{\rm opt}_2,\cdots,\eul^{\rm opt}_K\}$   such that for each $i\in\{1,\cdots,K\}$,
 \begin{eqnarray}\label{eq:greltup}
\nonumber {\rm Either}\; \eul^{\rm opt}_i=\hat{\eul}_i,\;\;{\rm or}\;\;\\
\eul^{\rm opt}_i\notin\hat{\Gulc}\;\&\;(\hat{\Gulc}\setminus\hat{\eul}_i)\cup\eul^{\rm opt}_i\in\Iulk.
 \end{eqnarray}
  Then, letting $\tilde{\Gulc}\define \hat{\Gulc}\setminus \Gulc^{\rm opt}$ we have the  chain of inequalities (\ref{eq:greltupN}) given on the top of the next page which yields the desired result.
 \begin{figure*} \begin{eqnarray}\label{eq:greltupN}
\begin{split}
 g(\hat{\Gulc},\alpha)=\sum_{i=1}^K\delta_i=\sum_{i:\hat{\eul}_i\in\hat{\Gulc}\cap\Gulc^{\rm opt}}(g(\hat{\Gulc}_{i-1}\cup\hat{\eul}_i,\alpha)-g(\hat{\Gulc}_{i-1},\alpha)) + \sum_{i:\hat{\eul}_i\in\tilde{\Gulc}}(g(\hat{\Gulc}_{i-1}\cup\hat{\eul}_i,\alpha)-g(\hat{\Gulc}_{i-1},\alpha))\\
 \geq \sum_{i:\hat{\eul}_i\in\hat{\Gulc}\cap\Gulc^{\rm opt}}(g(\hat{\Gulc}_{i-1}\cup\tilde{\Gulc}\cup\hat{\eul}_i,\alpha)-g(\hat{\Gulc}_{i-1}\cup\tilde{\Gulc},\alpha)) + \sum_{i:\hat{\eul}_i\in\tilde{\Gulc}}(g(\hat{\Gulc}_{i-1}\cup\hat{\eul}_i,\alpha)-g(\hat{\Gulc}_{i-1},\alpha))\\
 =g(\hat{\Gulc},\alpha)-g(\tilde{\Gulc},\alpha) + \sum_{i:\hat{\eul}_i\in\tilde{\Gulc}}(g(\hat{\Gulc}_{i-1}\cup\hat{\eul}_i,\alpha)-g(\hat{\Gulc}_{i-1},\alpha))\\
 \geq g(\hat{\Gulc},\alpha)-g(\tilde{\Gulc},\alpha) + \sum_{i:\eul^{\rm opt}_i\notin\hat{\Gulc}}(g(\hat{\Gulc}_{i-1}\cup\eul^{\rm opt}_i,\alpha)-g(\hat{\Gulc}_{i-1},\alpha))\\
 \geq g(\hat{\Gulc},\alpha)-g(\tilde{\Gulc},\alpha) + \sum_{i:\eul^{\rm opt}_i\notin\hat{\Gulc}}(g(\hat{\Gulc}\cup\eul^{\rm opt}_i,\alpha)-g(\hat{\Gulc},\alpha))\\
 \geq g(\hat{\Gulc},\alpha)-g(\tilde{\Gulc},\alpha) +  g(\Gulc^{\rm opt}\cup\hat{\Gulc},\alpha)-g(\hat{\Gulc},\alpha),
\end{split}
 \end{eqnarray}\end{figure*}
 In (\ref{eq:greltupN}) the first inequality follows from submodularity of $g(.,\alpha)$ and the fact that for each $i:\hat{\eul}_i\in\hat{\Gulc}\cap\Gulc^{\rm opt}$, $\hat{\Gulc}_{i-1}\subseteq\hat{\Gulc}_{i-1}\cup\tilde{\Gulc}$ and  $\hat{\eul}_i\notin\hat{\Gulc}_{i-1}\cup\tilde{\Gulc}$. The second inequality follows from (\ref{eq:greltup}) along with the fact that for each $i:\eul^{\rm opt}_i\notin\hat{\Gulc}$, the greedy algorithm would have considered
  $\eul^{\rm opt}_i$ but choose $\hat{\eul}_i$ instead since the latter offered a better (greater) change in system utility. The third inequality also follows from submodularity of $g(.,\alpha)$ and the fact that   each $i:\eul^{\rm opt}_i\notin\hat{\Gulc}$ we have $\hat{\Gulc}_{i-1}\subseteq\hat{\Gulc}$, and the  final inequality also follows from submodularity of $g(.,\alpha)$. Note that none of the steps require $g(.,\alpha)$ to be a non-negative set function or that the changes in system utility should be non-negative. The second relation in   (\ref{eq:gineq0}) can be proved in an analogous fashion.\endproof

\textbf{Proposition 2}:\\
For any given $\alpha$,  the greedy stage yields an output $\hat{\Gulc}$ such that
\begin{eqnarray}\label{eq:gnbds}
\begin{split}
& g(\hat{\Gulc},\alpha)\geq g(\Gulc^{\rm opt},\alpha)/2 \label{eq:gbd0} \;\;\forall\;\alpha \in (0,1),\\
 &g(\hat{\Gulc},\alpha)\geq  g(\Gulc^{\rm opt},\alpha) -2\ln(2) \label{eq:gbd1} \;\;\forall\;\alpha =1,\\
&(3-2^{\alpha})g(\hat{\Gulc},\alpha)\leq g(\Gulc^{\rm opt},\alpha)\;\;\forall\;\alpha>1. \label{eq:gbdbg1}
\end{split}
 \end{eqnarray}
\proof For $\alpha\in(0,1)$, since $g(.,\alpha)$ is  submodular and non-decreasing, we can readily obtain (\ref{eq:gbd0}) from (\ref{eq:gineq0}) by observing that $g(\Gulc^{\rm opt}\cup\hat{\Gulc},\alpha)\geq g(\Gulc^{\rm opt},\alpha)$
 and $g(\hat{\Gulc},\alpha)\geq g(\hat{\Gulc}\setminus\Gulc^{\rm opt},\alpha)$. Note that (\ref{eq:gbd0}) is the classical result derived earlier \cite{nemhaus:algo}.
 For $\alpha=1$, the result in (\ref{eq:gbd1}) is novel and thus more interesting. To prove (\ref{eq:gbd1}), we first re-write the bound in (\ref{eq:gineq0}) as
 \begin{eqnarray}\label{eq:gdinb1}
\begin{split}
g(\hat{\Gulc},1)\geq g(\Gulc^{\rm opt},1) + g(\Gulc^{\rm opt}\cup\hat{\Gulc},1)- g(\Gulc^{\rm opt},1) \\- g(\hat{\Gulc}\setminus\Gulc^{\rm opt},1).
\end{split}
 \end{eqnarray}
 Then, recall from (\ref{eq:setfb1}) that $g(.,1)$ is the sum of a modular function and a submodular function where the latter depends only on the user weights, and the sum of these weights across all users  is unity.
Consequently,  we can infer that
\begin{eqnarray}\label{eq:gdinb2}
\begin{split}
 &g(\Gulc^{\rm opt}\cup\hat{\Gulc},1)- g(\Gulc^{\rm opt},1) - g(\hat{\Gulc}\setminus\Gulc^{\rm opt},1) \\= &-\sum_b(x_b+y_b)\ln(x_b+y_b)+  \sum_b(z_b+y_b)\ln(z_b+y_b)\\ &+ \sum_b(x_b -z_b)\ln(x_b -z_b)
\end{split}
\end{eqnarray}
where $x_b$ is the sum of weights of users associated to TP $b$ by the greedy solution (and hence is known), $y_b+z_b$ is the sum of weights of users associated to TP $b$ by the optimal solution and
 $z_b$ is the sum of weights of users associated to TP $b$ by both the greedy and the optimal solutions. Note further that $\sum_b x_b=\sum_b(y_b+z_b)=1$.
 Combining (\ref{eq:gdinb2}) with (\ref{eq:gdinb1}) we can obtain the following specialized bound,
 \begin{eqnarray}\label{eq:gdinb3}
\begin{split}
 g(\hat{\Gulc},1)\geq  &g(\Gulc^{\rm opt},1) \\&+ \min_{y_b,z_b\geq 0;z_b\leq x_b\;\forall b\atop \sum_b (y_b+z_b)=1}\{ -\sum_b(x_b+y_b)\ln(x_b+y_b)\\&+\sum_b(z_b+y_b)\ln(z_b+y_b) \\&+ \sum_b(x_b -z_b)\ln(x_b -z_b)\}.
\end{split}
\end{eqnarray}
 Then, by using the  K.K.T. conditions for the optimization problem in the RHS of (\ref{eq:gdinb3}), it can be shown that the minima is attained at
  $y_b=x_b\;\&\;z_b=0\;\forall\;b$ so that
  \begin{eqnarray*}
\begin{split}
  &\min_{y_b,z_b\geq 0;z_b\leq x_b\;\forall b\atop \sum_b (y_b+z_b)=1}\{-\sum_b(x_b+y_b)\ln(x_b+y_b) \\&+  \sum_b(z_b+y_b)\ln(z_b+y_b) + \sum_b(x_b -z_b)\ln(x_b -z_b)\} = -2\ln(2).
\end{split}
\end{eqnarray*}
  This proves the result in (\ref{eq:gbd1}).
  Next, we consider $\alpha>1$ and specialize the bound in (\ref{eq:gineq0}) as
  \begin{multline}
g(\hat{\Gulc},\alpha)\leq g(\Gulc^{\rm opt},\alpha) + g(\Gulc^{\rm opt}\cup\hat{\Gulc},\alpha)- g(\Gulc^{\rm opt},\alpha)- g(\hat{\Gulc}\setminus\Gulc^{\rm opt},\alpha)\\
= g(\Gulc^{\rm opt},\alpha) +  \sum_b((v_b+t_b)^{\alpha} -(v_b+u_b)^{\alpha} \\- (t_b-u_b)^{\alpha}),
\end{multline}
  where now $t_b$ is the sum of gains of all users associated to TP $b$ by the greedy solution (i.e., sum of  $\Theta_k^{(b)}(\alpha)$ in (\ref{eq:thetabg1}) for all tuples in $\hat{\Gulc}\cap\Omega^{(b)}$ and hence is known) so that  $g(\hat{\Gulc},\alpha)=\sum_bt_b^\alpha$. $v_b+u_b$ is the sum of gains of all users associated to TP $b$ by the optimal solution and
 $u_b$ is the sum of gains of all users associated to TP $b$ by both the greedy and the optimal solutions.
  Clearly, then we can further bound
 \begin{eqnarray}\label{eq:gdinb5}
\begin{split}
g(\hat{\Gulc},\alpha)\leq g(\Gulc^{\rm opt},\alpha) +&\max_{v_b,u_b\geq 0; u_b\leq t_b\;\forall b\atop \sum_b (v_b+u_b)^{\alpha}\leq g(\hat{\Gulc},\alpha)}\{ \sum_b((v_b+t_b)^{\alpha} \\&-(v_b+u_b)^{\alpha} - (t_b-u_b)^{\alpha}) \}
\end{split}
\end{eqnarray}
 Again invoking the  K.K.T. conditions for the optimization problem in the RHS of (\ref{eq:gdinb5}), it can be shown that the maxima is attained at
  $v_b=t_b\;\&\;u_b=0\;\forall\;b$ so that
  \begin{eqnarray}\label{eq:gdinb4}
\begin{split}
  &\max_{v_b,u_b\geq 0; u_b\leq t_b\;\forall b\atop \sum_b (v_b+u_b)^{\alpha}\leq g(\hat{\Gulc},\alpha)}\{ \sum_b\left((v_b+t_b)^{\alpha} -(v_b+u_b)^{\alpha} - (t_b-u_b)^{\alpha}\right) \}\\&= (2^\alpha -2)g(\hat{\Gulc},\alpha)
\end{split}
\end{eqnarray}
  This then proves the result in (\ref{eq:gbdbg1}).\endproof

\textbf{Proposition 3}:\\
The GLS algorithm  for any given $\Delta\geq 0$  yields an output $\breve{\Gulc}$ such that
for any given $\alpha>1$
\begin{eqnarray*}
\begin{split}
&g(\Gulc^{\rm opt},\alpha)\geq g(\breve{\Gulc},\alpha)+ K{(1-\Delta)g(\breve{\Gulc},\alpha)}-h(\breve{\Gulc},\alpha)\\
\end{split}
 \end{eqnarray*}
 and for any given $\alpha\in(0,1)$
\begin{eqnarray*}
\begin{split}
&g(\Gulc^{\rm opt},\alpha)\leq g(\breve{\Gulc},\alpha)+ K{(1+\Delta)g(\breve{\Gulc},\alpha)}-h(\breve{\Gulc},\alpha).\\
\end{split}
 \end{eqnarray*}
Further, for $\alpha=1$
\begin{eqnarray}\label{eq:mfineq0A}
\nonumber g(\Gulc^{\rm opt},1)\leq \;\;\;\;\;\;\;\;\;\;\;\;\;\;\;\;\\
g(\breve{\Gulc},1)+ K{(1+\Delta \sgn(g(\breve{\Gulc},1)))g(\breve{\Gulc},1)}-h(\breve{\Gulc},1),
 \end{eqnarray}
where, $h(\breve{\Gulc},\alpha)=\sum_{n=1}^Kg(\breve{\Gulc}\setminus\breve{\eul}_n,\alpha)+\sum_{n=1}^K (g(\tilde{\Omegaul},\alpha)-g(\tilde{\Omegaul}\setminus\breve{\eul}_{n},\alpha))$,
 for any subset $\tilde{\Omegaul}\subseteq \Omegaul: \Gulc^{\rm opt}\cup \breve{\Gulc}\subseteq \tilde{\Omegaul}$.
\proof
We prove the result for $\alpha>1$ and the result for $\alpha$ in other regimes can be derived similarly.
 We  again invoke a result on maximal members in a matroid  \cite{lee:nmsub}, to deduce that without loss of generality, we can expand $\breve{\Gulc}=\{\breve{\eul}_1,\breve{\eul}_2,\cdots,\breve{\eul}_K\}$  and expand $\Gulc^{\rm opt}=\{\eul^{\rm opt}_1,\eul^{\rm opt}_2,\cdots,\eul^{\rm opt}_K\}$ 
 such that for some $m\in\{0,1,\cdots,K\}$,
 \begin{eqnarray}\label{eq:reltup}
\nonumber\eul^{\rm opt}_n=\breve{\eul}_n,\;\;\forall\;n\leq m\;\;\& \;\;\eul^{\rm opt}_n\neq\breve{\eul}_n,\;\;\forall\;n> m\\
(\breve{\Gulc}\setminus\breve{\eul}_n)\cup\eul^{\rm opt}_n\in\Iulk,\;\forall\; n:m+1\leq n\leq K.
 \end{eqnarray}
 Then, we have the following inequalities for each $n=m+1,\cdots,K$.
 \begin{eqnarray}\label{eq:mfineq1}
\begin{split}
  g(\breve{\Gulc}\cup\eul^{\rm opt}_n,\alpha)-g(\breve{\Gulc},\alpha)&\geq g((\breve{\Gulc}\setminus\breve{\eul}_n)\cup\eul^{\rm opt}_n,\alpha)-g(\breve{\Gulc}\setminus\breve{\eul}_n,\alpha)\\
 &\geq (1-\Delta)g(\breve{\Gulc},\alpha)-g(\breve{\Gulc}\setminus\breve{\eul}_n,\alpha)
\end{split}
   \end{eqnarray}
 where the first inequality follows from the supermodularity of $g(.,\alpha)$ and the second one  follows from the local swap optimality of $\breve{\Gulc}$, i.e.,
 \begin{eqnarray}\label{eq:fineq1b}
  g((\breve{\Gulc}\setminus\breve{\eul}_n)\cup\eul^{\rm opt}_n,\alpha)-g(\breve{\Gulc},\alpha)\geq -\Delta g(\breve{\Gulc},\alpha).
 \end{eqnarray}
  Thus, we have that
 \begin{eqnarray}\label{eq:mfineq2}
\begin{split}
 & \sum_{n=m+1}^K (g(\breve{\Gulc}\cup\eul^{\rm opt}_n,\alpha)-g(\breve{\Gulc},\alpha))\\ &\geq  \sum_{n=m+1}^K((1-\Delta)g(\breve{\Gulc},\alpha)- g(\breve{\Gulc}\setminus\breve{\eul}_n,\alpha))
\end{split}
   \end{eqnarray}
and   due to the supermodularity of $g(.,\alpha)$,
\begin{eqnarray}\label{eq:mfineq3}
\begin{split}
 &\sum_{n=m+1}^K g(\breve{\Gulc}\cup\eul^{\rm opt}_n,\alpha)-g(\breve{\Gulc},\alpha)\\&\leq \sum_{n=m+1}^K (g(\breve{\Gulc}\cup\{\eul^{\rm opt}_{m+1},\cdots,\eul^{\rm opt}_n\},\alpha)\\&-g(\breve{\Gulc}\cup\{\eul^{\rm opt}_{m+1},\cdots,\eul^{\rm opt}_{n-1}\},\alpha)\\
  &= g(\breve{\Gulc}\cup\Gulc^{\rm opt},\alpha)- g(\breve{\Gulc},\alpha).
\end{split}
   \end{eqnarray}
Next, we have the bound
\begin{eqnarray}\label{eq:mfineq3b}
\begin{split}
&g(\breve{\Gulc}\cup\Gulc^{\rm opt},\alpha)\\&=g(\Gulc^{\rm opt},\alpha) + \sum_{n=m+1}^K (g(\Gulc^{\rm opt}\cup\{\breve{\eul}_{m+1},\cdots,\breve{\eul}_n\},\alpha)\\&-g(\Gulc^{\rm opt}\cup\{\breve{\eul}_{m+1},\cdots,\breve{\eul}_{n-1}\},\alpha)\\
   &\leq g(\Gulc^{\rm opt},\alpha) + \sum_{n=m+1}^K (g(\tilde{\Omegaul},\alpha)-g(\tilde{\Omegaul}\setminus\breve{\eul}_{n},\alpha)),
\end{split}
   \end{eqnarray}
   for any subset $\tilde{\Omegaul}\subseteq \Omegaul: \Gulc^{\rm opt}\cup \breve{\Gulc}\subseteq \tilde{\Omegaul}$.
Combining the bounds in (\ref{eq:mfineq1}), (\ref{eq:mfineq3}) and (\ref{eq:mfineq3b}) we get
\begin{multline}\label{eq:mfineq4}
g(\Gulc^{\rm opt},\alpha)\geq  g(\breve{\Gulc},\alpha)+ \sum_{n=m+1}^K((1-\Delta)g(\breve{\Gulc},\alpha)- g(\breve{\Gulc}\setminus\breve{\eul}_n,\alpha))\\ - \sum_{n=m+1}^K (g(\tilde{\Omegaul},\alpha)-g(\tilde{\Omegaul}\setminus\breve{\eul}_{n},\alpha)).
 \end{multline}
The RHS of (\ref{eq:mfineq4}) is further lower bounded to obtain the desired result in (\ref{eq:mfineq0A}), by extending the summation from $1$ to $K$, where we note that each term
 $((1-\Delta)g(\breve{\Gulc},\alpha)- g(\breve{\Gulc}\setminus\breve{\eul}_n,\alpha)) - (g(\tilde{\Omegaul},\alpha)-g(\tilde{\Omegaul}\setminus\breve{\eul}_{n},\alpha))\leq 0$ since $\Delta\geq 0$ and $g(.,\alpha)$ is supermodular and non-negative. \endproof


\textbf{Proposition 4}:\\

For non-negative non-decreasing submodular set functions, which we recall does not hold for our set functions when $\alpha\geq 1$, a somewhat lesser known result is that a restricted version of the greedy algorithm can also yield identical constant factor approximation \cite{goundan:sub}. We next establish a similar result with respect to the bounds in Lemma 1 and Proposition  \ref{prop:bd11}. In particular, we first detail the  restricted   greedy algorithm in Table \ref{algo:rglb}.
\begin{table}
\caption{{\bf  Restricted Greedy Algorithm}}\label{algo:rglb}
\begin{algorithmic}[1]
\STATE Initialize with any ordering $\pi(.)$ defined on $\Uc$ and $\hat{\Gulc}^{\rm rg}=\phi$.
\STATE \textbf{For} $k=1\;{\rm to}\;K$,
\STATE Determine $(\pi(k),b')$ as the tuple in $\Omegaul$ which offers the best change among all tuples $(\pi(k),b)\in\Omegaul$.
\STATE Update $\hat{\Gulc}^{\rm rg}=\hat{\Gulc}^{\rm rg}\cup(\pi(k),b')$.
 \STATE   \textbf{End For}.
\STATE Output  $\hat{\Gulc}^{\rm rg}$.
\end{algorithmic}\vspace{-.5cm}
\end{table}
Next, we show that for any given ordering $\pi(.)$,  the restricted greedy algorithm yields a solution that also satisfies the bounds in Lemma 1 for all $\alpha$. Thus, the solution of the restricted greedy algorithm also satisfies the bounds in Proposition \ref{prop:bd11} for all $\alpha$ and hence yields the same firm guarantees for all $\alpha \in \left(0,\frac{\ln(3)}{\ln(2)}\right)$.
Towards this end, we expand the solution yielded by the restricted greedy algorithm as $\hat{\Gulc}^{\rm rg}=\{\hat{\eul}^{{\rm rg},\pi}_{1},\hat{\eul}^{{\rm rg},\pi}_{2},\cdots,\hat{\eul}^{{\rm rg},\pi}_{K}\}$ where $\hat{\eul}^{{\rm rg},\pi}_{i}$ denotes the tuple added at the $i^{th}$  step as per the ordering $\pi(.)$. Then, notice that all the arguments in the proof of Lemma 1 go through even upon replacing
 $\hat{\Gulc}$ with $\hat{\Gulc}^{\rm rg}$ and $\hat{\eul}_i$ with $\hat{\eul}^{{\rm rg},\pi}_{i},\;\forall\;i$. The key point to note here is that we do not require the changes in system utility obtained across the steps to be ordered. In other words, we do not use the fact that these changes obtained during the greedy stage of the GLS algorithm are ordered as $\delta_1\geq\delta_2\geq\cdots\geq\delta_K$ when $\alpha\leq 1$ or as  $\delta_1\geq\delta_2\geq\cdots\geq\delta_K$ when $\alpha > 1$, whereas no such ordering is ensured for those
obtained during the restricted greedy algorithm.

Notice that the the aforementioned result applies to any ordering $\pi(.)$. We will exploit this fact  along with a result that the solution yielded by the distributed greedy algorithm maps exactly to that yielded by the restricted greedy algorithm for a particular ordering. We will suppose that $\alpha\leq 1$ since the arguments we make directly extend to the case where $\alpha>1$.
Let $\hat{\eul}^{\rm dg}_{1},\cdots, \hat{\eul}^{\rm dg}_{K}$ be the tuples selected by the distributed greedy algorithm, where we assume that tuples $\hat{\eul}^{\rm dg}_{1},\cdots, \hat{\eul}^{\rm dg}_{m1}$ are selected in the first window, tuples $\hat{\eul}^{\rm dg}_{m1+1},\cdots, \hat{\eul}^{\rm dg}_{m2}$ are selected in the second window and so on. Moreover, let $u_1,u_2,\cdots,u_{m1}$ denote the  corresponding users in the tuples selected in the first window, let $u_{m1+1},u_{m1+2},\cdots,u_{m2}$ denote the  corresponding users in the tuples selected in the second window and so on. We define an ordering $\pi(.)$
 such that $\pi(k)=u_k,\;k=1,\cdots,K$. Note here that we can pick any arbitrary order to list the users (tuples) selected by the distributed greedy algorithm within each  window.
We will show that
\begin{eqnarray}\label{eq:Desres}
\hat{\eul}^{{\rm rg},\pi}_{k}=\hat{\eul}^{\rm dg}_{k},\;\forall\; k=1,\cdots,K
\end{eqnarray}
 which proves the desired result.
Consider the tuples selected in the first window. Each user $u_i\;i=1,\cdots,m1$ chooses the TP yielding the best change  in system utility assuming zero current load on all TPs.  Thus, it is readily seen that $\hat{\eul}^{{\rm rg},\pi}_{1}=\hat{\eul}^{\rm dg}_{1}$. Consider
 the TP choice of user $u_i,\;i=2,\cdots,m1$ made as $\hat{\eul}^{\rm dg}_{i}=\arg\max_{(u_i,b),b\in\Bc}\{g((u_i,b),\alpha)\}$. By sub-modularity   of $g(.,\alpha)$ for $\alpha\leq 1$  and the fact that the TPs chosen by the admitted users in each window are all distinct, we have that
 \begin{eqnarray}\label{eq:dsch}
\nonumber \hat{\eul}^{\rm dg}_{i}=\arg\max_{(u_i,b),b\in\Bc}\left\{g(\{\hat{\eul}^{\rm dg}_{1}\cup\cdots\cup\hat{\eul}^{\rm dg}_{i-1}\}\cup(u_i,b),\alpha)-\right.\\\left.g(\{\hat{\eul}^{\rm dg}_{1}\cup\cdots\cup\hat{\eul}^{\rm dg}_{i-1}\},\alpha)\right\}.\;
 \end{eqnarray}
 Put differently, given that tuples  $\{\hat{\eul}^{\rm dg}_{1}\cup\cdots\cup\hat{\eul}^{\rm dg}_{i-1}\}$ have been already chosen, the best TP for user $u_i$ will still be the one    in $\hat{\eul}^{\rm dg}_{i}$. This is because upon selecting the tuples $\{\hat{\eul}^{\rm dg}_{1}\cup\cdots\cup\hat{\eul}^{\rm dg}_{i-1}\}$ the loads of the TPs in these tuples will increase, whereas that of the one in $\hat{\eul}^{\rm dg}_{i}$ will remain unchanged. Thus, the system utility change obtained if user $u_i$ joined each one of those TPs (given these selections) will be inferior, respectively, to what that user assumed when making its decision (since it used a lower value of the load).
On the other hand, the   system utility change obtained if user $u_i$ joined the TP    in $\hat{\eul}^{\rm dg}_{i}$ (given that tuples  $\{\hat{\eul}^{\rm dg}_{1}\cup\cdots\cup\hat{\eul}^{\rm dg}_{i-1}\}$ have been already selected) will be identical to what it assumed. 
 Then, from (\ref{eq:dsch}) we have that
 $\hat{\eul}^{{\rm rg},\pi}_{i}=\hat{\eul}^{\rm dg}_{i},\;\forall\; i=1,\cdots,m1$.
 The same argument applies to each subsequent window upon observing that all users that are selected in that window use  load values that account for all associations made in all prior windows. 
Thus, we can conclude that (\ref{eq:Desres}) is true which proves our claim for the distributed greedy algorithm.

In this context, we note that another distributed greedy algorithm can be obtained by altering the TP-side procedure to one where  in each window  each TP   admits only the user offering the best change among all users that have requested it in that window. From the proof detailed above, it can be verified that this variant also yields identical performance guarantees.

\textbf{Distributed LS Stage}:\\

We will show that this
{\em  distributed LS stage provably converges and the solution it yields upon convergence yields the same guarantees in Proposition \ref{prop:ls2}.}

To prove this claim, we define a {\em system state}  to be a feasible user association, i.e., an association where each user is associated to one TP. Thus, the set of all possible system states is finite and comprises of all feasible user associations. Let us define a system state to be an {\em absorbing state} if at that state, for each user the switch yielding the best change in system utility  (\ref{eq:LSfeaspair}) does 
not yield a relative improvement better than $\Delta$ (cf. (\ref{eq:LSDpair1}) and (\ref{eq:LSDpair2})). Clearly, the optimal system state (which yields the globally optimal system utility) is an absorbing state so that the set of absorbing states is finite and non-empty.
Further, given any non-absorbing state it can be verified that we can construct a finite sequence of states that begins at the given state and ends at an absorbing one, such that each transition from any state to the next one in that sequence involves a migration of exactly one user and yields a relative improvement (in the system utility) better than $\Delta$.

Next, considering the distributed LS algorithm, it is readily seen that the broadcast   of the current load information at the start of each window corresponds to a system state.  Moreover, without loss of generality, we can assume that each user  which sends a request in any window is accepted with a strictly positive probability
that  depends only on the system state at the begining of that window and the user index.
Consequently, the sequence of states seen across the broadcast slots forms an {\em absorbing, time homogeneous Markov Chain}. 
Hence, convergence to an absorbing state is guaranteed. Indeed, the expected number of steps for convergence can be obtained from the analysis in \cite{zhang:dwr}.
Finally, since the bound in Proposition \ref{prop:ls2} is satisfied by any absorbing state, we can assert the claimed guarantee for the distributed LS algorithm is true.

\textbf{AF Optimization}

We first discuss a  distributed implementation that ensures  no loss in performance. Towards this end, it is readily seen that for any fixed activation vector $\rhob$ the optimization over $\bf{s,g}$ decouples into smaller problems which can be separately solved at each TP.
We notice, however, that the AF variables in the GP formulation in (\ref{fixbetagreater1}) induce coupling constraints. Nevertheless, this issue can be addressed by exploiting a useful decomposition technique from \cite{palomar:jsac} and introducing local copies for the AF  variables. 
 In particular,  for each AF variable $\rho_b$, we  introduce $B-1$ local copies $\rho_{b',b},\;\forall\;b'\in\Bc:b'\neq b$ ($\rho_{b',b}$ is the copy of $\rho_b$ maintained at TP $b'$) and re-write the GP in (\ref{fixbetagreater1}) including these local copies along with  equality constraints $\rho_b=\rho_{b',b},  \;\forall\;b'\in\Bc:b'\neq b, \forall\;b\in\Bc$, as the following.
\begin{equation}
\begin{split}
	&\mbox{min}_{\{\rho_b,\{\rho_{b,b'}\}\},{\bf z},{\bf t} }\{\sum_{b\in{B}}{z_{b}^\alpha}\}\\
	&\mbox{subject to}\\
	 &\sum_{k\in\Uc^{(b)}}{z_{b}^{-1} \tilde{w}_kt_{k,b}^{1/\alpha-1}}\leq{1}\ \ \forall  \;b\in\Bc\\
	 &\frac{ t_{k,b}\rho_{b}^{-1}+\mathbb{E}[s_{k,b}(\betab_k)e_{k,b}(\betab_k,\rho_b,\{\rho_{b,b'}\})]}{1+\mathbb{E}[\log(s_{k,b}(\betab_k))]}\leq1, \forall\; k\in\Uc^{(b)},b\in\Bc\\
&\rho_{b'}=\rho_{b,b'},\;\forall\;b'\neq b\;\&\;b\in\Bc.
\label{fixbetagreater1A}
\end{split}
	\end{equation}
 The problem in (\ref{fixbetagreater1B}) can be decomposed into smaller sub-problems by using a Lagrange multiplier for each equality constraint (a.k.a. consistency price variable). However, to ensure that the sub-problems are also convex,
we first adopt the (usual) change of variables $\tilde{z}_b=\ln(z_b)$, $\tilde{t}_{k,b}=\ln(t_{k,b}),\;\forall\;k\in\Uc^{(b)}$,  $\tilde{\rho}_b=\ln(\rho_b)$
 and $\tilde{\rho}_{b,b'}=\ln(\rho_{b,b'}),\;\forall\;b'\neq b$, for all $b\in\Bc$.
Then, we note that the equality constraints can be written as $\tilde{\rho}_{b'}=\tilde{\rho}_{b,b'}$ forall $b'\neq b\;\&\;b\in\Bc$. This transformed problem is presented below
\begin{equation}
\begin{split}
	&\mbox{min}_{\{\tilde{\rho}_b,\{\tilde{\rho}_{b,b'}\}\},{\bf \tilde{z}},{\bf \tilde{t}} }\sum_{b\in{B}}{\exp(\alpha\tilde{z}_{b})}\\
	&\mbox{subject to}\\
	 &\ln(\sum_{k\in\Uc^{(b)}} {\tilde{w}_k\exp(-\tilde{z}_{b}+(1/\alpha-1)\tilde{t}_{k,b})} )\leq{0}\ \ \forall  \;b\in\Bc\\
	 &\ln\left(\frac{ \exp(\tilde{t}_{k,b}-\tilde{\rho}_{b})+\mathbb{E}[s_{k,b}(\betab_k)\tilde{e}_{k,b}(\betab_k,\tilde{\rho}_b,\{\tilde{\rho}_{b,b'}\})]}{1+\mathbb{E}[\log(s_{k,b}(\betab_k))]}\right)\leq 0, \forall\; k,b\\
&\tilde{\rho}_{b'}=\tilde{\rho}_{b,b'},\;\forall\;b'\neq b\;\&\;b\in\Bc.
\label{fixbetagreater1B}
\end{split}
	\end{equation}
where we use $\tilde{e}_{k,b}(.,.)$ to denote the MSE as function of the transformed variables.
Note that (\ref{fixbetagreater1B})
 a convex optimization problem with its utility function (decoupled across TPs) and where the constraints   are either also decoupled or are coupled linear equality ones. Thus, a  decomposition technique introduced in \cite{palomar:jsac} is now directly applicable and accordingly we introduce a Lagrange multiplier for each equality constraint constraint.
Each TP $b$ can then  separately solve a convex sub-problem  and the multipliers can be updated using the sub-gradient method in a distributed manner \cite{palomar:jsac}.

\subsection{$\alpha=1$} 
	  AF optimization problem over the set of variables $\boldsymbol{\rho}=\{\rho_{b}\}\; \forall b\in{B}$  in $\alpha=1$ regime is given by
\begin{eqnarray}
	\mbox{maximize}_{{\boldsymbol{\rho}}\in{\bf{[0,1]}} } & \left\{\sum_{b\in{B}}{\sum_{k\in{\Uc^{(b)}}}{ w_k\ln(R_{k,b}(\rhob))}}\right\}
\label{origbetaequal1}
	\end{eqnarray}
	
	The problem of interest is equivalent to
        \begin{eqnarray}
	\mbox{minimize}_{{\boldsymbol{\rho}}\in{\bf{[0,1]}} } & \left\{\sum_{b\in{B}}{\sum_{k\in{\Uc^{(b)}}}{ w_k\ln(\frac{1}{R_{k,b}(\rhob)})}}\right\}
\label{eqorigbetaequal1}
	\end{eqnarray}

     As done in case of $\alpha>1$, we reduce (\ref{eqorigbetaequal1}) and fix $\bf{s,g}$ to obtain
       \begin{equation}
\begin{split}
        &\mbox{min}_{\boldsymbol{\rho}\in{\bf{[0,1]}},{\bf t}\geq{\bf 0} }
\left\{\sum_{b\in{B}}\sum_{k\in{\Uc^{(b)}}}{ w_k\ln(t_{k,b})^{-1}}\right\}\\
        &\mbox{subject to}\\
        &\frac{ t_{k,b}\rho_{b}^{-1}+\mathbb{E}(s_{k,b}(\betab_k)e_{k,b}(\betab_k,\rhob))}{1+\mathbb{E}(\log(s_{k,b}(\betab_k)))}\leq1\ \ \forall{b,k}
\label{genbetaequal1}
\end{split}
\end{equation}
	 We consider change of variables $t_{k,b}=\exp({\tilde{t}_{k,b}})\;\forall{b}\in{B}, k\in{\Uc^{(b)}}$ and $\rho_{b}=\exp({\tilde{\rho}_{b}}) \ \forall b\in{B}$. Let $a_{k,b}=\frac{1}{1+\mathbb{E}(\log{s_{k,b}(\betab_k))}}$.
Now (\ref{genbetaequal1}) can be further reduced to
 \begin{equation}
\begin{split}
 &\mbox{min}_{\tilde{\boldsymbol{\rho}}\leq{\bf{0}},\tilde{\boldsymbol{t}}} \left\{\sum_{b\in{B}}{\sum_{k\in{\Uc^{(b)}}}{ -w_k\tilde{t}_{k,b}}}\right\}\\
 &\mbox{subject to}\\
&\log(a_{k,b}\exp{(-\tilde{\rho}_b+\tilde{t}_{k,b})}\\
&+a_{k,b}\mathbb{E}(s_{k,b}(\betab_k)(\left|g_{k,b}(\betab_k)\sqrt{\beta_{k,b}}-1\right|^2+\left|g_{k,b}(\betab_k)\right|^2)) \\
&  + \sum_{b'\neq b}\exp(\tilde{\rho}_{b'})a_{k,b}\mathbb{E}(s_{k,b}(\betab_k)\left|g_{k,b}(\betab_k)\right|^2{\beta_{k,b'}}))\leq{0} 
\label{fixbetaequal1}
\end{split}
\end{equation}
Note that (\ref{fixbetaequal1}) is a convex optimization problem. Again, we use alternating optimization approach to obtain the solution of (\ref{origbetaequal1}).
We use solution of (\ref{altrate}) to minimize over $\boldsymbol{s,g}$ when  $\boldsymbol{\rho}$ is fixed  and further use (\ref{fixbetaequal1}) to minimize over $\boldsymbol{\rho}$ when $\boldsymbol{s,g}$ are fixed.
\subsection{$\alpha<1$} 
  AF optimization problem over the set of variables $\boldsymbol{\rho}=\{\rho_{b}\}\; \forall b\in{B}$  in $\alpha\in{(0,1)}$ regime is given by

 \begin{eqnarray}
	\mbox{max}_{{\boldsymbol{\rho}}\in{\bf{[0,1]}} } &\left\{ \sum_{b\in{B}}{(\sum_{k\in{\Uc^{(b)}}}{ \tilde{w}_k(R_{k,b}(\rhob))^{1/\alpha-1}})^\alpha}\right\}
\label{origbetaless1}
	\end{eqnarray}

	Where $\tilde{w}_k=(\frac{w_k}{1-\alpha})^{1/\alpha}$. We choose $C=\sum_{b\in{B}}{(\sum_{k\in{\Uc^{(b)}}}{ \tilde{w}_k(\mathbb{E}(\log(1+\beta_{k,b})))^{1/\alpha-1}})^\alpha}$. Now we
 use the reduction for (\ref{origbetaless1}) as done in (\ref{altrate})-(\ref{eqgenbetagreater1}) and further fix $\bf{s}$ and $\bf{g}$. We obtain the following optimization problem in variables
$\boldsymbol{\rho},{\bf z},{\bf t}$
\begin{equation}
\begin{split}
	&\mbox{min}_{{\boldsymbol{\rho}}\in{\bf{[0,1]}},{\bf z}\geq{\bf 0},{\bf t}\geq{\bf 0} }\ \ C-\sum_{b\in{B}}{z_{b}^\alpha}\\
	&\mbox{subject to} \\
 &z_{b}\leq{\sum_{k\in{\Uc^{(b)}}}{ \tilde{w}_kt_{k,b}^{1/\alpha-1}}}  \ \ \forall{b} \\
	    &\frac{ t_{k,b}\rho_{b}^{-1}+\mathbb{E}(s_{k,b}(\betab_k)e_{k,b}(\betab_k,\rhob))}{1+\mathbb{E}(\log(s_{k,b}(\betab_k)))}\leq1\ \ \forall{b,k}
\label{fixbetaless1}
	\end{split}
\end{equation}
Adding an extra variable $y$, the above problem (\ref{fixbetaless1}) is equivalent to
\begin{equation}
\begin{split}
	&\mbox{minimize}_{{y}{\geq0},{\boldsymbol{\rho}}\in{\bf{[0,1]}},{\bf z}\geq{\bf 0},{\bf t}\geq{\bf 0} }\ \ \left\{y\right\}\\
	&\mbox{subject to} \\
          &{\frac{C}{y+\sum_{b\in{B}}{z_{b}^\alpha}}\leq1}\\
           &\frac{z_{b}}{{\sum_{k\in{\Uc^{(b)}}}{ \tilde{w}_kt_{k,b}^{1/\alpha-1}}}}\leq1  \forall{b} \\
	   &\frac{ t_{k,b}\rho_{b}^{-1}+\mathbb{E}(s_{k,b}(\betab_k)e_{k,b}(\betab_k,\rhob))}{1+\mathbb{E}(\log(s_{k,b}(\betab_k)))}\leq1\ \ \forall{b,k}
\label{eqfixbetaless1}
\end{split}
\end{equation}
To transform this optimization problem (\ref{eqfixbetaless1}) into a GP, we need to apply the single condensation method \cite{book:chiangGP} on the first two constraints of (\ref{eqfixbetaless1}), which are of the form
of ratio of a monomial and a posynomial.
Let $\boldsymbol{X}=(y,\boldsymbol{z})$ and $\boldsymbol{t}=\{t_{k,b}\}\ \ \forall b\in{B},\forall{k}\in{\Uc^{(b)}}$. For any current $\tilde{\bf{X}},\tilde{\bf{t}}$
we define
\begin{eqnarray}
\tilde{f}(\boldsymbol{X})=(\frac{yf(\tilde{\boldsymbol{X}})}{\tilde{y}})^{\frac{\tilde{y}}{f(\tilde{\boldsymbol{X}})}}\prod_{b}{(\frac{z_{b}^{\alpha}f(\tilde{\boldsymbol{X}})}{\tilde{z}_{b}^{\alpha}})^{\frac{\tilde{z}_{b}^{\alpha}}{f(\tilde{\boldsymbol{X}})}}}
\end{eqnarray}
Where $f(\tilde{\boldsymbol{X}})=\tilde{y}+\sum_{b\in{B}}{\tilde{z}_{b}^\alpha}$. We also define
\begin{eqnarray}
\tilde{h}_{b}(\boldsymbol{t})=\prod_{{k}\in{\Uc^{(b)}}}{(\frac{t_{k,b}^{1/\alpha-1}h_{b}(\tilde{\boldsymbol{t}})}{\tilde{t}_{k,b}^{1/\alpha-1}})^{\frac{\tilde{t}_{k,b}^{1/\alpha-1} \tilde{w}_k}{h_{b}(\tilde{\boldsymbol{t}})}}}
\end{eqnarray}
Where $h_{b}(\tilde{\boldsymbol{t}})=\sum_{{k}\in{\Uc^{(b)}}}{ \tilde{w}_k\tilde{t}_{k,b}^{1/\alpha-1}}$.
Then the following approximate problem is a GP
\begin{equation}
\begin{split}
	&\mbox{minimize}_{{\boldsymbol{X}}{\geq0},{\boldsymbol{\rho}}\in{\boldsymbol{[0,1]}},{\boldsymbol t}\geq{\boldsymbol 0} }\left\{y\right\}\\
	&\mbox{subject to}\\
          &{\frac{C}{\tilde{f}(\boldsymbol{X})}\leq1}\\
           &\frac{z_{b}}{\tilde{h}_{b}(\boldsymbol{t})}\leq1  \ \ \forall{b}\\
	    &\frac{ t_{k,b}\rho_{b}^{-1}+\mathbb{E}(s_{k,b}(\betab_k)e_{k,b}(\betab_k,\rhob))}{1+\mathbb{E}(\log(s_{k,b}(\betab_k)))}\leq1\ \ \forall{b,k}
\label{approxfixbetaless1}
\end{split}
\end{equation}
%

\begin{normalsize}

 \bibliography{DTXWup}
\bibliographystyle{ieeetr}

\end{normalsize}

%
%
%
%
%
%
%
%



\end{document}

%% file: Associationv2wbnE.tex
\input{abbreviations2bb}
\setlength\arraycolsep{1pt}
\newtheorem{definition}{Definition}
\newtheorem{corollary}{Corollary}
\newtheorem{remark}{Remark}
\newtheorem{theorem}{Theorem}
\newtheorem{lemma}{Lemma}
\newtheorem{proposition}{Proposition}
\newtheorem{prof}{Proof}
\newtheorem{condition}{Condition}
\newtheorem{observation}{Observation}
\def\IR{{\mathbb R}}
\def\IC{{\mathbb C}}
\newcommand{\test}{\mbox{$
\begin{array}{c}
\stackrel{
\stackrel{\textstyle H_1}{\textstyle >}
}
{
\stackrel{\textstyle <}{\textstyle H_0}
}
\end{array}
$}}
\begin{abstract}
We consider the problem of maximizing the alpha-fairness utility over the downlink of a heterogeneous wireless network (HetNet) by jointly optimizing {\em the association} of users to transmission points (TPs) and the {\em activation fractions} of all TPs.
Activation fraction of each TP is the fraction of the frame duration for which it is active,  and together these fractions influence  the interference seen in the network.
To address this joint optimization problem we adopt an  approach wherein the activation fractions and the user associations are optimized in an alternating manner. The sub-problem of determining the optimal activation fractions is solved using an  auxiliary function method that we show is provably convergent and is amenable to distributed implementation. On the other hand, the sub-problem of determining the user association is solved via a simple combinatorial algorithm. Meaningful performance guarantees are derived and  a distributed variant offering  identical  guarantees is also proposed. The significant benefits of using the proposed algorithms are then  demonstrated via realistic simulations. 
\end{abstract}
%
%
%
\section{Introduction}

It is well established by now that future cellular   networks will be dense HetNets formed by a multitude of disparate transmission points  deployed in a highly irregular fashion \cite{standard:3gpp_TR36872}. In a majority of these   deployments, the transmission points (TPs) will be connected to each other by a  non-ideal backhaul with a relatively high latency (several dozens of milliseconds). An unfortunate consequence of such a high latency is that it renders unsuitable 
 resource management (RM) schemes that   strive to coordinate and obtain allocation   decisions within a fine time-scale (e.g., 1 ms in LTE HetNets)   \cite{gesbert:binary,Wei:INFO2011,Lozano:IA,Huang:MBC,Sanjabi:joint,tajer:robust}. 
Instead, semi-static resource  management schemes where  RM    is performed  at two time scales,  are  better suited since they are  more robust towards backhaul latency. Broadly speaking, in any such semi-static scheme the RM that is done at a coarse  {\em frame} level granularity (that is at-least as large as the backhaul latency) entails coordination among TPs  based   on averaged (not instantaneous) slowly varying metrics. 
On the other hand, the RM in such a scheme   that is done at a much finer {\em slot} level  granularity involves no coordination among TPs  and is done independently by each active TP based on fast changing metrics   \cite{koshy:twoTS,veciana:LB,madan:Rlb,yuw:UA,prasad:joint}.
The semi-static scheme that we propose in this paper decides at the onset of each frame 
 {\em which set of users should each TP   serve over that frame such that each user is served by exactly one TP (user association) and  how often  should each TP transmit over that frame (activation fraction of that TP)}.

The problem at hand is quite challenging  due to the well recognized interference coupling problem. Indeed,  while increasing the activation fraction (AF) of a TP will help it serve more users (or serve a given set of users better), it injects more interference  to all users being served by other TPs. User Association (without AF optimization) is by itself a popular HetNet RM scheme, wherein the interference coupling problem is simplified by assuming that  the interference that would be seen by any user upon being associated to any TP remains static. Association is then determined by  optimizing a system utility  \cite{ye:lb,Buram:genpf,li:mrpf,Son:LB,bethan:arxiv},  or by minimizing a cost function  given traffic demands \cite{veciana:LB}, or by adopting a game theoretic framework \cite{ehsan:rat}. Joint optimization of user association along with other system resources, such as power and bandwidth in the downlink  \cite{madan:Rlb,yuw:UA,Sanjabi:joint,prasad:joint,prasad:twoTS} and user powers and TP locations in the uplink \cite{Yates:ULp,Altman:Loc}, has also received significant attention. Considering the downlink which is our focus in this paper, we see that  the alternating optimization framework is a popular approach to ensure tractability, and that  binary (on-off) power control  has been found to be particularly effective in terms of being robust and capturing most of the available gains with a small signalling footprint.   The latter observation has led to another  promising downlink semi-static RM technique that is fully compliant with the LTE standard, and seeks  to capture the benefits of slot-level coordinated binary power control  over a HetNet with a non ideal backhaul. This scheme   combines user association with partial muting of the high power Macro TP, i.e.,  the  Macro TP is allowed to be active (or  transmit with a pre-determined power) for any fraction of the total number of slots in a frame. The choice of this AF for the macro TP is optimized together with the user association   \cite{bedekar:wiopt,hanly:eicic}.  The macro TP then adopts a muting pattern (which includes its on-off status on all the slots) conforming to the determined AF. Notice that  the exact on-off status of the macro TP on all the slots is not optimized. Indeed,   doing so can   be detrimental since coordination done at a coarse time-scale based on the available averaged metrics cannot adapt to the fast changing channel and interference conditions seen across the slots.


Recent studies have shown that topologies without one common dominant interferer will be ubiquitous  and in such cases optimizing the AF of only one TP is not enough. The problem
we seek to solve is geared exactly towards such deployments. 
One attempt to solve our problem would be to extend the solutions proposed for the aforementioned scheme, but  then it becomes immediately clear that those solutions do not scale when activation fractions for all TPs have to be optimized. This is because those solutions explicitly maintain a rate for each TP-user link under each possible interference pattern, which grow exponentially in the number of TPs.  In this paper, we propose a simple formulation that imposes activation fractions 
and yields one average rate expression for each TP-user link. The latter expression is conservative and is a closed-form function of all activation fractions. 
Interestingly, in the absence of fast fading our rate expression reduces to the approximate rate expression introduced  in
\cite{siomina:lb} (see also \cite{fehske:load}),  which considered   the problem of determining activation fractions  to meet a given set of user traffic demands for a given user association. 
We confirm the observation made in those works  that the rate expression is in-fact quite accurate over practical HetNets.
Our main contributions are as follows:
\vspace{-.05cm}
\setlength\itemindent{0pt}
\begin{itemize}[leftmargin=*]
  \setlength{\parskip}{0pt}
  \setlength{\itemsep}{0pt plus 1pt}
    \item We adopt $\alpha-$fairness utility as the system wide utility which generalizes all popular utility functions \cite{linshroff:infocom}, wherein we also allow for assigning any arbitrary set of weights (reflecting priorities) to the users. We develop centralized and distributed   algorithms that yield  good solutions for any given   fairness parameter $\alpha$. These algorithms are obtained by adopting an 
alternating optimization based approach. The latter approach is well justified since the problem at hand is intractable and our goal is to obtain unified low-complexity algorithms that are suitable for all $\alpha$. 
\item For the discrete user-association sub-problem, we first prove that this sub-problem itself is NP-hard and proceed to completely characterize the underlying set function that needs to be optimized. We then suggest and comprehensively analyze a simple centralized combinatorial algorithm (referred to as the GLS algorithm) that   involves a Greedy stage followed by Local Search improvements.
Our analysis yields meaningful and novel readily computable performance guarantees for all $\alpha$. 
Previous related works have considered the proportional fairness (PF) utility and proposed combinatorial user association algorithms \cite{li:mrpf,prasad:joint}. Our results when specialized to the case of the weighted PF utility (by setting  $\alpha=1$)  reveal that GLS is optimal up-to a constant additive factor of $-2\ln(2)$. Thus,   a simple algorithm yields optimality up-to an additive constant factor, a fact that was hitherto only established for a significantly more complex algorithm \cite{li:mrpf} (whose run-time can depend on the input weights). Upon further specializing to the case with identical user weights, we see that the guarantee proved for a greedy algorithm  in \cite{prasad:joint} has an instance dependent (non constant) additive factor. Interestingly,   our simulation results indicate that in this special case the association yielded by GLS is identical to the optimal one obtained via  another more complex algorithm from \cite{prasad:joint}.
\item We derive a distributed version of the GLS algorithm and prove that {\em remarkably} it provides guarantees identical to its centralized counterpart.
This distributed version  
 requires network assistance in the form of periodic broadcast of system load information similar to that proposed earlier in \cite{deb:mota}. 
The main novelty of our approach is that we are able to configure each user to consider the system  utility gain in contrast to the  selfish gain used in the user-centric approach adopted by \cite{deb:mota,ehsan:rat} and more recently in \cite{bethan:arxiv}. Consequently,  we can establish  guarantees (with respect to the optimal system utility) and provable convergence for our distributed algorithms for all $\alpha$. We note here that convergence of the user-centric approach to a Nash equilibrium was proved in \cite{ehsan:rat} for particular choices of $\alpha$ and the recent and independent work in \cite{bethan:arxiv} has identified conditions under which the Nash equilibrium is (near-)optimal.  
\item For the continuous AF optimization sub-problem  we adopt the auxiliary function method and show that it is provably convergent. Such a method has been used for precoder optimization originally over the single-cell downlink in \cite{rajiv_agarwal} and over the multi-cell downlink in \cite{tajer:robust} followed by \cite{yuw:UA,Sanjabi:joint}. %
We note however that  unlike those works we incorporate  fading coefficients that change at two different time scales. Further,
a key step 
 in our case entails a novel GP formulation, which we show can also be implemented in a distributed manner. 
\item Finally, the performance of all our   algorithms is compared to appropriate baselines via extensive simulations over a HetNet topology generated as per 3GPP LTE guidelines.  
Our results highlight the significant gains that can be achieved in realistic HetNet deployments via the joint optimization.
\end{itemize}

\section{Problem Statement}\label{sec:PS}
Considering the downlink in a HetNet, let $\Uc=\{1,\cdots,K\}$ denote the set of users  and  let $\Bc$ denote the set of transmission points (TPs) with cardinality $|\Bc|=B$.
Further,
 suppose that the time axis is divided into multiple frames, where each frame  consists of several consecutive slots.
The fast fading coefficients for each user are assumed to change across slots in an independent identically distributed (i.i.d.) manner, while   the slow fading coefficients  are assumed to change across frames in an i.i.d. manner.
 {\em The choice of the activation fraction for each TP along with the user association for all TPs is made once for each frame to optimize the system utility}. This choice can be based on the slow fading realization  in that frame but does not consider any previous such choices. Each   TP then independently implements its   per-slot scheduling policy  over the users associated with it in that frame, where the latter scheduling policy respects the assigned activation fraction and can exploit the instantaneous fast fading coefficients seen by the associated users on each slot. Consequently, we can suppress the dependence on the frame and slot indices in the following.

In order to
 formulate an optimization problem for determining the user association and activation fractions, we derive an  average rate that each user can obtain over a frame of interest, under any given user association and activation fractions.
Towards this end, let $\Uc^{(b)},\;\forall b\in\Bc$ denote any given set of users associated to TP $b$ over the frame  and let $\rhob=[\rho_b]_{b\in\Bc}$ denote the activation vector, where $\rho_b\in[0,1]$ denotes the activation fraction assigned to TP $b$.
 We proceed by  assuming that each TP $b$  allocates a fraction $\gamma_{k,b}\in [0,1]$ of the frame to serve each associated user $k\in\Uc^{(b)}$, such that $\sum_{k\in\Uc^{(b)}}\gamma_{k,b}=1$, where these fractions are determined at the onset of the frame. In particular,  
  each TP is assumed to adopt an optimal  fractional round robin per-slot scheduling policy. Note that an efficient per-slot scheduling policy (cf. \cite{stol:grad}) that can adapt to the instantaneous fading and interference conditions seen across all the slots, will be at-least as good (in terms of optimizing the given utility). 
Next, we assume that the activation fraction of each TP $b$ is implemented via a Bernoulli  random variable $\Xc_b$ with $E[\Xc_b]=\rho_b$, that is i.i.d. across slots in the frame and is independent of all other random variables. Specifically,  TP $b$ is assumed to transmit (with a fixed  power) when $\Xc_b=1$ and remain silent otherwise.
Then, an average rate   that can be achieved for user $k\in\Uc^{(b)}$ is given by,
\begin{eqnarray}\label{eq:avrateI}
	\gamma_{k,b}\rho_b\mathbb{E}\left[\log\left(1+\frac{\beta_{k,b}}{1+\sum_{b'\neq b}{\beta_{k,b'}\Xc_{b'}}}\right)\right]
	\end{eqnarray}
where the the desired channel gain $\beta_{k,b}$ and the interfering channel gains $\{\beta_{k,b'}\}$ are random variables that  include both fast and slow fading as well as noise normalized transmit powers, and the expectation is over the activation variables as well as the fast fading. Upon invoking the fact that the instantaneous  rate in (\ref{eq:avrateI}) is convex in the activation variables, which we recall are independent of the   fast fading coefficients, we can further lower bound (\ref{eq:avrateI}) to obtain 
\begin{eqnarray}\label{eq:avrateI2}
	r_k=\gamma_{k,b}\underbrace{\rho_b\mathbb{E}\left[\log\left(1+\frac{\beta_{k,b}}{1+\sum_{b'\neq b}{\beta_{k,b'}\rho_{b'}}}\right)\right]}_{\define R_{k,b}(\rhob)},
	\end{eqnarray}
where now the expectation  is over only  the fast fading. Note that $r_k$ in (\ref{eq:avrateI2}) depends on the slow fading realization (comprising of the path losses and shadowing factors)  over the frame of interest.
Letting $\rb=[r_1,\cdots,r_K]$ denote the vector of such  conservative  rates obtained for all the $K$ users over the frame, the achieved system utility is given by
\begin{eqnarray}
\sum_{k\in\Uc}w_k u(r_k,\alpha),
\end{eqnarray}
where $\alpha> 0$ is a tunable fairness parameter
and
\begin{eqnarray}\label{eq:peruser}
u(r_k,\alpha) = \left\{
\begin{array}{rl}
 \frac{r_k^{(1-\alpha)}}{1-\alpha}   &  \;\;\;\;\;\;   \alpha\in (0,1) \\
  \log(r_k) & \;\;\;\;\;\;\;\;\alpha=1 \\
   -\frac{r_k^{(1-\alpha)}}{\alpha-1}  & \;\;\;\;\;\;\alpha>1
\end{array}
\right.
\end{eqnarray}
 and $w_k> 0$ denotes the weight of user $k\in\Uc$. These weights can be used to assign different   priorities to different users and   we assume that they are normalized, i.e.,
  $\sum_{k\in\Uc}w_k=1$.
 We can now write our  problem, which is a mixed optimization problem, as
  \begin{equation}\boxed{\begin{aligned}\label{eq:BJigoriginal}
   \max_{\rhob\in[0,1]^B;x_{k,b}\in\{0,1\}; \atop\gamma_{k,b}\in[0,1]\;\forall\;k,b}\left\{\sum_{k\in\Uc}\sum_{b\in\Bc}x_{k,b}\left( w_k u(\gamma_{k,b}R_{k,b}(\rhob))\right)\right\}\\
 {\rm s.t. } \sum_{b\in\Bc}x_{k,b}= 1,\;\forall\;k\in\Uc;\;
 \sum_{k\in\Uc}\gamma_{k,b}= 1\;\forall\;b\in\Bc.
\end{aligned}}\end{equation}
Note that in (\ref{eq:BJigoriginal}) the binary  variable $x_{k,b}$ is one if user $k$ is associated to TP $b$ and zero otherwise, so that the first set of constraints   ensures that each user is associated with only one TP. Consequently,  $\Uc^{(b)}\define\{k:x_{k,b}=1\}_{k\in\Uc}$ yields the user set associated with TP $b$. 
Note that in (\ref{eq:BJigoriginal}),  we enforce $\{\Uc^{(b)}\}_{b\in\Bc}$ to be a partition of $\Uc$. This is meaningful and indeed important since we are targeting short-term optimality by maximizing a system utility independently over each frame. 
The joint optimization problem in (\ref{eq:BJigoriginal}) is unfortunately intractable. 
Consequently, we  develop 
an alternating optimization framework to solve the joint problem in (\ref{eq:BJigoriginal}). We will demonstrate that although the user association and activation fractions are optimized assuming conservative rates and optimal fractional round robin per-slot scheduling policies at all TPs, the obtained solution retains its  significant gains  even without these assumptions.
To improve readability the proofs of all the following propositions are deferred to  the appendix.
\vspace{-\topsep}
\section{User Association}
\label{UA}
We adopt the convention that $0\ln(0)=0$ and consider any fixed activation vector $\rhob$ with strictly positive elements (otherwise any TP $b$ with $\rho_b=0$ can be simply removed). We proceed to systematically consider the user-association sub-problem of (\ref{eq:BJigoriginal})  given by
  \begin{equation}\begin{split}\label{eq:Bigoriginal}
  & \mbox{max}_{x_{k,b}\in\{0,1\}; \atop\gamma_{k,b}\in[0,1]\;\forall\;k,b}\left\{\sum_{k\in\Uc}\sum_{b\in\Bc}x_{k,b}\left( w_k u(\gamma_{k,b}R_{k,b}(\rhob))\right)\right\}\\
 &{\rm s.t. } \sum_{b\in\Bc}x_{k,b}= 1,\;\forall\;k\in\Uc;\;
 \sum_{k\in\Uc}\gamma_{k,b}= 1\;\forall\;b\in\Bc,
 \end{split}\end{equation}%
  over three regimes defined by the values of $\alpha$. 
  We first define a ground set, $\Omegaul=\{(k,b):k\in\Uc,b\in\Bc\}$, that consists of all possible tuples and where each tuple $(k,b)$ denotes an association of user $k$ to TP $b$. 
Then, we also define the set $\Omegaul^{(b)}=\{(k,b):k\in\Uc\}$ for each TP $b\in\Bc$ which consists of all tuples whose TP is $b$, along with the set $\Omegaul_{(k)}=\{(k,b):b\in\Bc\}$ for each user $k$ which consists of all tuples whose user is $k$.
 Finally, we define a family of sets
 $\Iulk$, as the one which includes each subset of $\Omegaul$ such that the tuples in that subset have mutually distinct users. Formally,
 \begin{eqnarray}
 \Gulc\subseteq\Omegaul: |\Gulc\cap\Omegaul_{(k)}|\leq 1\;\forall\; k \Leftrightarrow \Gulc\in\Iulk.
 \end{eqnarray}
We start with the regime $\alpha>1$ 
 and note that for any given user association, i.e, for any given feasible choice of variables $\{x_{k,b}\}$,  (\ref{eq:Bigoriginal}) is a continuous optimization problem. Moreover, it is separable across the set of TPs and for each TP $b\in\Bc$, we have a convex optimization problem over the set of variables $\{\gamma_{k,b}\}$ for $k\in\Uc:x_{k,b}=1$.  Using K.K.T. conditions it is verified in the appendix that for each TP $b\in\Bc$
 \begin{eqnarray}\label{eq:tporiginal}
\begin{split}
  &\mbox{max}_{\gamma_{k,b}\in[0,1]\;\forall\;k\atop\sum_{k\in\Uc}\gamma_{k,b}= 1}\left\{\sum_{k\in\Uc}x_{k,b}\left( w_k u(\gamma_{k,b}R_{k,b}(\rhob) )\right)\right\}
 = \\&- \left(\sum_{k\in\Uc}x_{k,b}\left(w_k \frac{(R_{k,b}(\rhob) )^{1-\alpha}}{\alpha-1}\right)^{1/\alpha}\right)^{\alpha}
\end{split}
\end{eqnarray}
Consequently, upon defining
\begin{eqnarray}\label{eq:thetabg1}
\begin{split}
\nonumber &\Theta_k^{(b)}(\alpha)=\left(w_k \frac{(R_{k,b}(\rhob) )^{1-\alpha}}{\alpha-1}\right)^{1/\alpha},\;\forall\;\alpha>1,
\end{split}
\end{eqnarray}
 (\ref{eq:Bigoriginal}) reduces  to the following discrete optimization problem.
\begin{eqnarray}\label{eq:bg1}
\begin{split}
&\mbox{min}_{x_{k,b}\in\{0,1\}\;\forall\;k,b\atop\sum_{b\in\Bc}x_{k,b}=1\;\forall\;k }\left\{\sum_{b\in\Bc} \left(\sum_{k\in\Uc}x_{k,b}\Theta_k^{(b)}(\alpha)\right)^{\alpha}\right\}.
\end{split}
\end{eqnarray}
Considering the case $\alpha\in (0,1)$,   (\ref{eq:Bigoriginal}) reduces to
\begin{eqnarray}\label{eq:bg2}
\begin{split}
&\mbox{max}_{x_{k,b}\in\{0,1\}\;\forall\;k,b\atop\sum_{b\in\Bc}x_{k,b}=1\;\forall\;k }\left\{\sum_{b\in\Bc} \left(\sum_{k\in\Uc}x_{k,b}\Theta_k^{(b)}(\alpha)\right)^{\alpha}\right\},
\end{split}
\end{eqnarray}

  where $\Theta_k^{(b)}(\alpha)=\left(w_k \frac{(R_{k,b}(\rhob) )^{1-\alpha}}{1-\alpha}\right)^{1/\alpha},\;\forall\;\alpha\in (0,1)$.

Recalling the sets $\Omegaul,\Omegaul_{(k)},\Omegaul^{(b)}$ defined before, we further define the set function $g:2^{\Omegaul}\to\Reals$ as
 \begin{eqnarray}\label{eq:setfbgl1}
\begin{split}
& g(\Gulc,\alpha)= \sum_{b\in\Bc}{(\sum_{(k',b')\in\Gulc\cap\Omega^{(b)}}{\Theta_{k'}^{(b')}(\alpha)})}^{\alpha},
\end{split}
 \end{eqnarray}
 $\forall\;\Gulc\subseteq\Omegaul,\Gulc\neq\phi$ with $g(\phi,\alpha)=0$, where $\phi$ denotes the empty set.
 The minimization problem in (\ref{eq:bg1}) is now re-formulated as
 \begin{eqnarray}\label{eq:bg1r}
\begin{split}
 &\mbox{min}_{\Gulc: \Gulc\in\Iulk\;\&\; |\Gulc|=K} \{g(\Gulc,\alpha)\},
\end{split}
 \end{eqnarray}
 whereas the maximization problem in (\ref{eq:bg2}) can be re-formulated as
 \begin{eqnarray}\label{eq:bg2r}
\begin{split}
 &\mbox{max}_{\Gulc: \Gulc\in\Iulk\;\&\; |\Gulc|=K} \{g(\Gulc,\alpha)\}.
\end{split}
 \end{eqnarray}
 Similarly, for $\alpha=1$, (\ref{eq:Bigoriginal}) can be reformulated as in (\ref{eq:bg2r}) but where $g(\phi,1)=0$ and   for all $\Gulc\subseteq\Omegaul:\Gulc\neq\phi$ 
 \begin{eqnarray}\label{eq:setfb1}
\begin{split}
   \mbox{g}(\Gulc,1)=   &\sum_{(k,b)\in\Gulc}w_k\ln(w_{k} R_{k,b}(\rhob))-\\
  &\sum_{b\in\Bc}(\sum_{(k',b')\in\Gulc\cap\Omega^{(b)}}w_{k'})\ln\left(\sum_{(k',b')\in\Gulc\cap\Omega^{(b)}}w_{k'}\right).
\end{split}
 \end{eqnarray}
  We offer the following result. 
 \begin{proposition}\label{prop:bn1}
 For any $\alpha>0$, the user association sub-problem in (\ref{eq:Bigoriginal}) is NP-hard. Further,
 for any $\alpha>1$, the set function $g(.,\alpha)$ is a normalized, non-negative and non-decreasing supermodular set function. For any $\alpha\in (0,1)$, the set function $g(.,\alpha)$ is a normalized, non-negative and non-decreasing submodular set function.
 The set function $g(.,1)$ is a normalized submodular set function.
 \end{proposition}
Note that the set function $g(.,1)$ need not be non-negative nor non-decreasing.
\subsection{GLS: A Unified Algorithm}\label{subsec:UA}
In Table \ref{algo:lb} we propose the GLS Algorithm, which is a simple combinatorial algorithm to solve the problem in (\ref{eq:Bigoriginal}). It considers the respective re-formulated versions in (\ref{eq:bg1r}) or (\ref{eq:bg2r}) and comprises of two stages.
  The first one is the greedy stage (steps 1 to 6).
Here in each greedy iteration the feasible tuple $(k',b')$ (with respect to  the ones already selected so far) offering the  best change in system utility is selected, until no such tuple can be found. In particular,
   $(k',b')$ is determined      as
 \begin{eqnarray*}
\begin{array}{rl}
   &\arg\max_{(k,b)\in\Omegaul:\hat{\Gulc}\cup(k,b)\in\Iulk}\{g(\hat{\Gulc}\cup(k,b),\alpha)- g(\hat{\Gulc},\alpha)\}, \alpha\leq 1,\\
    &\arg\min_{(k,b)\in\Omegaul:\hat{\Gulc}\cup(k,b)\in\Iulk}\{g(\hat{\Gulc}\cup(k,b),\alpha)- g(\hat{\Gulc},\alpha)\},\alpha>1
   \end{array}
 \end{eqnarray*}

The second stage of GLS is  {\em local search improvement}   and comprises of steps 7 to 13. Here,
a feasible pair of tuples is determined in each local search iteration as $(k',b_1),(k',b_2)=$
\begin{eqnarray}\label{eq:LSfeaspair}
 \left\{
\begin{array}{rl}
   \arg\max_{k\in\Uc\;\&\;b,b'\in \Bc\atop (k,b)\in\breve{\Gulc},(k,b')\notin\breve{\Gulc}}\{ g(\breve{\Gulc}\cup (k,b')\setminus (k,b),\alpha)\}, & \alpha\leq 1,\\
    \arg\min_{k\in\Uc\;\&\;b,b'\in \Bc\atop (k,b)\in\breve{\Gulc},(k,b')\notin\breve{\Gulc}} \{g(\breve{\Gulc}\cup (k,b')\setminus (k,b),\alpha)\}, & \alpha>1
   \end{array}\;\;
\right.
 \end{eqnarray}
 and the corresponding relative improvement is deemed to be better than $\Delta$ by checking if
 \begin{multline}\label{eq:LSDpair1}
  g((\breve{\Gulc}\cup (k',b_2)\setminus (k',b_1)),\alpha)-g(\breve{\Gulc},\alpha)>\\\Delta \sgn(g(\breve{\Gulc},\alpha))g(\breve{\Gulc},\alpha),\;\alpha\leq 1,
 \end{multline}
\begin{multline}\label{eq:LSDpair2}
  g((\breve{\Gulc}\cup (k',b_2)\setminus (k',b_1)),\alpha)-g(\breve{\Gulc},\alpha)<\\-\Delta g(\breve{\Gulc},\alpha),\;\alpha>1,
 \end{multline}
 where $\sgn(x)=1,\;\forall x\geq 0$ and $-1$ otherwise.
\begin{table}
\caption{{\bf GLS Algorithm}}\label{algo:lb}
\begin{small}
\begin{algorithmic}[1]
\STATE Initialize with $\alpha$, $\Delta\geq 0$, ${\rm MaxIter}\geq 1$, $\hat{\Gulc}=\phi$ and   $\Uc'=\Uc$.
\STATE \textbf{Repeat}
\STATE Determine $(k',b')$ as the tuple in $\Omegaul$ which offers the best change among all tuples $(k,b)\in\Omegaul$ such that $\hat{\Gulc}\cup(k,b)\in\Iulk$.
\STATE Update $\hat{\Gulc}=\hat{\Gulc}\cup(k',b')$ and $\Uc'=\Uc'\setminus \{k'\}$
 \STATE   \textbf{Until} $\Uc'=\phi$.
\STATE Set $\breve{\Gulc}=\hat{\Gulc}$, ${\rm Iter=0}$.
\STATE \textbf{Repeat}
\STATE Increment ${\rm Iter}={\rm Iter}+1$.
\STATE Find a pair of tuples: $(k',b_1)\in\breve{\Gulc}$ and $(k',b_2)\in\Omegaul\setminus\breve{\Gulc}$ such  that the relative improvement upon swapping $(k',b_1)\in\breve{\Gulc}$ with $(k',b_2)$ is better than $\Delta$.
   \STATE \textbf{If} such a pair exists then
\STATE Update $\breve{\Gulc}=\breve{\Gulc}\cup(k',b_2)\setminus (k',b_1)$.
\STATE \textbf{End If}
 \STATE   \textbf{Until} no such pair exists or ${\rm Iter}={\rm MaxIter}$.
\STATE Output  $\breve{\Gulc}$.
\end{algorithmic}\vspace{-.4cm}
\end{small}
\end{table}
We now proceed to analyze the performance of  GLS. We seek to bound the gap (by obtaining  easily computable bounds) between the optimal system utility   and the one returned by   GLS. Towards this end, let $\Gulc^{\rm opt}$ denote the optimal solution to the problem in (\ref{eq:bg1r}) for $\alpha>1$ or (\ref{eq:bg2r}) for $\alpha\in(0,1]$, and let $\breve{\Gulc},\hat{\Gulc}$ denote the   counterparts obtained by our algorithm as the final output and after the greedy stage, respectively.
  We will first analyze the performance of the greedy first stage. The challenge here is that the underlying set function need not be submodular (when $\alpha>1$) or it need not be non-negative and non-decreasing (when $\alpha=1$), which precludes us from directly applying the analysis  in \cite{nemhaus:algo,lee:nmsub}. To overcome this limitation, we   first derive  new bounds that relate the optimal solution to that returned by the greedy stage. These bounds are in-fact applicable to arbitrary  submodular or  supermodular set functions. 
We then specialize those bounds to the set functions of interest to us in (\ref{eq:setfbgl1}) and (\ref{eq:setfb1}) to obtain the following result.
 \begin{proposition}\label{prop:bd11}
For any given $\alpha$,  the greedy stage yields an output $\hat{\Gulc}$ such that
\begin{eqnarray*}\label{eq:gnbds}
\begin{split}
 &g(\hat{\Gulc},\alpha)\geq g(\Gulc^{\rm opt},\alpha)/2 \label{eq:gbd0} \;\;\forall\;\alpha \in (0,1),\\
& g(\hat{\Gulc},1)\geq  g(\Gulc^{\rm opt},1) -2\ln(2), \label{eq:gbd1} \\
&(3-2^{\alpha})g(\hat{\Gulc},\alpha)\leq g(\Gulc^{\rm opt},\alpha)\;\;\forall\;\alpha>1. \label{eq:gbdbg1}
\end{split}
 \end{eqnarray*}
\end{proposition}
\begin{remark}
 Note that the last bound in Proposition \ref{prop:bd11}  is meaningful in the regime $\alpha\in\left(1,\frac{\ln(3)}{\ln(2)}\right)$ since then $3-2^\alpha>0$. As a result,
 we can deduce that for all $\alpha \in \left(0,\frac{\ln(3)}{\ln(2)}\right)$ the greedy stage of   GLS  itself provides firm (instance independent) guarantees.  However, as $\alpha$ is  increased, the performance of the greedy stage degrades compared to the optimal and the   local search stage of   GLS   becomes increasingly important. 
\end{remark}

We now proceed to examine the performance of the local search stage.  
 We leverage the techniques developed in \cite{lee:nmsub} to analyze the behaviour of a local search based algorithm when the latter is used to maximize non-negative submodular functions. Here, we extend those techniques to
 arbitrary submodular and non-negative supemodular functions and also obtain sharper bounds. We let $\eul=(k,b)$ denote any tuple in $\Omegaul$ and expand $\breve{\Gulc}$ as $\breve{\Gulc}=\{\breve{\eul}_1,\cdots,\breve{\eul}_K\}$.
\begin{proposition}\label{prop:ls2}
The GLS algorithm  for any given $\Delta\geq 0$  yields an output $\breve{\Gulc}$ such that
for any given $\alpha>1$
\begin{eqnarray*}\label{eq:mfineq0}
\begin{split}
&g(\Gulc^{\rm opt},\alpha)\geq g(\breve{\Gulc},\alpha)+ K{(1-\Delta)g(\breve{\Gulc},\alpha)}-h(\breve{\Gulc},\alpha)\\
\end{split}
 \end{eqnarray*}
 and for any given $\alpha\in(0,1)$
\begin{eqnarray*}\label{eq:mfineq0}
\begin{split}
&g(\Gulc^{\rm opt},\alpha)\leq g(\breve{\Gulc},\alpha)+ K{(1+\Delta)g(\breve{\Gulc},\alpha)}-h(\breve{\Gulc},\alpha).\\
\end{split}
 \end{eqnarray*}
Further, for $\alpha=1$
\begin{eqnarray*}\label{eq:mfineq0}
\nonumber g(\Gulc^{\rm opt},1)\leq \;\;\;\;\;\;\;\; \;\;\;\;\;\;\;\; \;\;\;\;\;\;\;\;\\
g(\breve{\Gulc},1)+ K{(1+\Delta \sgn(g(\breve{\Gulc},1)))g(\breve{\Gulc},1)}-h(\breve{\Gulc},1).
 \end{eqnarray*}
where, $h(\breve{\Gulc},\alpha)=\sum_{n=1}^Kg(\breve{\Gulc}\setminus\breve{\eul}_n,\alpha)+\sum_{n=1}^K (g(\tilde{\Omegaul},\alpha)-g(\tilde{\Omegaul}\setminus\breve{\eul}_{n},\alpha))$,
 for any subset $\tilde{\Omegaul}\subseteq \Omegaul: \Gulc^{\rm opt}\cup \breve{\Gulc}\subseteq \tilde{\Omegaul}$.
\end{proposition}
Finally, we note that one obvious choice of the subset $\tilde{\Omegaul}$ needed in Proposition  \ref{prop:ls2}  is $\tilde{\Omegaul}=\Omegaul$. However, for $\alpha>1$ this choice results in loose bounds and a better option is to set $\tilde{\Omegaul}$ to be the set obtained after removing each tuple $\eul$ satisfying $g(\eul,\alpha)>g(\breve{\Gulc},\alpha)$
 from $\Omegaul$. Note that no such tuple can be either in $\breve{\Gulc}$ or $\Gulc^{\rm opt}$.
Note then that the bounds in Propositions \ref{prop:ls2}   are easily computable once we have the output $\breve{\Gulc}$.

Regarding the complexity of   GLS, it is easy to see that the complexity of the greedy  stage is $O(K^2B)$. 
Moreover, each iteration in the local search (LS) stage
has $O(BK)$ complexity. 
Further,  simulation results presented later reveal that even for a large-sized HetNet ($KB\approx 3000$) only very few LS iterations (6 or less) are needed to capture the available gains.

\subsection{Distributed Version}\label{sec:dist}
The GLS algorithm presented above assumes a centralized implementation. While this assumption is not very restrictive due to the fact that the implementation is done at a coarse time scale relying on average (not instantaneous) estimates, in practice a distributed implementation brings its own advantages.
Remarkably, as we show next, for any given an activation vector $\rhob$, {\em a distributed variant of the GLS algorithm that offers identical performance guarantees is indeed possible.}
We make a (justifiable) assumption that  each user $k\in\Uc$
is supposed to know its weight $w_k$ and its (single-user) rates $R_{k,b}(\rhob),\;\forall\;b\in\Bc$. Consequently, each user $k$ can be configured to compute $\Theta^{(b)}_k(\alpha),\;\forall\;b\in\Bc$ given the fairness parameter $\alpha$. 
  $\Theta^{(b)}_k(\alpha),\;\forall\;k,b$ was defined before for all $\alpha\neq 1$ and here for later use
 we define $\Theta^{(b)}_k(1)=w_k,\;\forall\;k,b$.
We will first derive a distributed version of the greedy stage of the GLS algorithm. Recall that in this stage a feasible subset of tuples
 $\hat{\Gulc}$ is built up. Then, we note the {\em simple but key fact} that given any subset of selected tuples $\hat{\Gulc}\in\Iulk$, the change in system utility
 upon adding a tuple $(k,b)\notin\hat{\Gulc}$ to $\hat{\Gulc}$, given by $g(\hat{\Gulc}\cup (k,b),\alpha)- g(\hat{\Gulc},\alpha)$,  can be expressed as
 \begin{eqnarray*}
 \left\{
\begin{array}{rl}
 \Theta^{(b)}_k(1)\ln(\Theta^{(b)}_k(1)R_{k,b}(\rhob)) + \Psi^{(b)}(1)\ln(\Psi^{(b)}(1))\\- (\Theta^{(b)}_k(1)+\Psi^{(b)}(1))\ln(\Theta^{(b)}_k(1)+ \Psi^{(b)}(1)), \;&\; \alpha=1,\\
    (\Theta^{(b)}_k(\alpha)+\Psi^{(b)}(\alpha))^{\alpha}-(\Psi^{(b)}(\alpha))^{\alpha},\;\;\; &\;\;\;\; {\rm Else},\\
       \end{array}
\right.
 \end{eqnarray*}
where we define
$\Psi^{(b)}(\alpha)=\sum_{(k',b')\in\hat{\Gulc}\cap\Omegaul^{(b)}}\Theta^{(b')}_{k'}(\alpha),\;\forall\;\alpha$. 
As a result, each user $k$ (that has not associated to any TP yet) can compute the change in system  utility if it joins  any TP $b\in\Bc$, provided it knows $\Psi^{(b)}(\alpha)$, which we refer to as the
current {\em load} on TP $b$. This suggests a natural distributed algorithm (outlined in Table \ref{algo:distgs} as the distributed greedy stage) comprising of two parts, namely, the TP-side and the user-side procedures.
Considering the TP-side procedure, all TPs periodically broadcast their current load at the start of each time window on a designated slot, where the window size is chosen to accommodate all propagation, acknowledgement and processing delays, and where the broadcasting parameters (powers, assigned codes etc.) ensure that the loads can be reliably decoded by the users. We assume a particularly simple procedure where each TP admits only the first user (who has requested to associate) in each window.
Moving to the user-side procedure, each user uses the current loads to determine the TP yielding the best system utility change, where the best change corresponds to the largest change for $\alpha\leq 1$ and to the smallest change for $\alpha>1$. Note here that in each window (defined as the time interval between two consecutive load-broadcast slots) multiple associations can be done. Indeed, in each window, each TP that receives one or more user requests will admit one user, and each un-associated user will send one request. Hence, the distributed greedy stage will complete all associations in no more than $K$ windows.
We offer the following important result.

\begin{table}
\begin{small}
\caption{Distributed Greedy Stage}\label{algo:distgs}
\begin{algorithmic}[H]
\STATE TP-side procedure: At each TP $b\in\Bc$
\STATE \textbf{Repeat}
\STATE  \emph{Broadcast step}:
\STATE \hspace{0.5cm}Transmit {\em current load} $\Psi^{(b)}(\alpha)$
\STATE  {\em Monitoring Step}:
\STATE \hspace{0.5cm}\textbf{If}  {request from any user $k$ detected}
\STATE \hspace{0.8cm}\textbf{If}{ another user already admitted}
\STATE \hspace{1cm}Send NACK to the requesting user $k$
\STATE \hspace{0.8cm}\textbf{Else}
\STATE \hspace{1cm}Admit user $k$ and send an ACK
\STATE \hspace{1cm}Update current load $\Psi^{(b)}(\alpha)\to \Psi^{(b)}(\alpha)+\Theta^{(b)}_{k}(\alpha)$
\STATE  \hspace{0.8cm}\textbf{EndIf}
\STATE \hspace{0.5cm}\textbf{EndIf}
\STATE \textbf{Until} No user request and no other TP changes its load
\STATE User-side procedure: At each user $k\in\Uc$
\STATE \textbf{Repeat}
\STATE   \emph{Listening step}:
\STATE \hspace{0.5cm}Decode all {\em current loads} $\Psi^{(b)}(\alpha),\;\forall\;b\in\Bc$
\STATE {\em Request Step}:
\STATE  \hspace{0.5cm}Evaluate  utility change upon joining each TP $b\in\Bc$
\STATE  \hspace{0.5cm}Determine TP $\hat{b}$ corresponding to best change
\STATE  \hspace{0.5cm}Send a request to associate to TP $\hat{b}$ along with $\Theta_k^{\hat{b}}(\alpha)$
\STATE \textbf{Until} ACK received from requested TP
\end{algorithmic}
\end{small}\vspace{-.4cm}
\end{table}
%
 \begin{proposition}\label{prop:rgb}
The solution obtained after the distributed greedy stage   yields the same guarantees as in Proposition \ref{prop:bd11}. 
\end{proposition}

We now consider the LS stage of the GLS algorithm and offer its distributed counterpart. This distributed algorithm is initiated once the (build-up) greedy stage terminates after associating each user to a TP.
 All TPs periodically broadcast their current load information at the start of each window on a designated slot. The load information of TP $b$ includes $\Psi^{(b)}(\alpha)$ as before. In addition, when $\alpha=1$ it also includes the term $\sum w_k\ln(w_kR_{k,b}(\rhob))$, where the sum is over all users currently associated to TP $b$. 
The first  key observation behind this algorithm is that given all the current load information, each user can determine its switch or migration that yields the best change in system utility (\ref{eq:LSfeaspair}). Moreover, it 
can also assess (via (\ref{eq:LSDpair1}) and (\ref{eq:LSDpair2}))   if that switch yields a relative improvement better than $\Delta$. 
Note here that in each window in order to ensure a distributed implementation we permit multiple users  to migrate, albeit to distinct TPs. Prima facie it is not apparent that the procedure will converge, since each user which migrates in any window only guarantees an improvement in system utility if no other  user migrates in that window.
The other key aspect which ensures convergence is the introduction of a {\em randomized decision rule} at each TP. This rule is described next and it is essential to ensure convergence to a solution at which  no  migration that yields a relative improvement better than $\Delta$ can be found. In particular, under this randomized rule, each TP $b$ that receives a request from some user $k$ sets its decision to accept to be negative if it has already admitted another user in that window. 
On the other hand, if no user has been admitted by it,
 that TP generates a binary-valued ($\{0,1\}$) random variable with a specified probability $p\in(0,1)$. It then sets its decision to be positive if the generated variable has value one, failing which it sets the decision to be negative.

\begin{table}
\begin{small}
\caption{Distributed LS Stage}\label{algo:distls}
\begin{algorithmic}[H]
\STATE TP-side procedure: At each TP $b\in\Bc$
\STATE \textbf{Repeat}
\STATE   \emph{Broadcast step}:
\STATE \hspace{0.5cm}Transmit {\em current load information} 
\STATE   {\em Monitoring Step}:
\STATE \hspace{0.5cm} \textbf{If}  {request to associate from any user $k$ detected}
\STATE \hspace{0.68cm}Determine decision via randomized rule
\STATE \hspace{0.68cm}\textbf{If}  {decision to accept is positive}
\STATE \hspace{0.8cm}Send ACK to user $k$
\STATE \hspace{0.8cm}Update current load information 
\STATE \hspace{0.68cm}\textbf{Else}
\STATE \hspace{0.8cm}Send NACK to user $k$
\STATE  \hspace{0.68cm}\textbf{EndIf}
\STATE  \hspace{0.5cm}\textbf{EndIf}
\STATE  \hspace{0.5cm}\textbf{If} {request to release from any user $k$ detected}
\STATE  \hspace{0.8cm}Release user $k$
\STATE  \hspace{0.8cm}Update current load information 
\STATE   \hspace{0.5cm}\textbf{EndIf}
\STATE \textbf{Until}  Convergence
\STATE User-side procedure: At each user $k\in\Uc$
\STATE \textbf{Repeat}
\STATE   \emph{Listening step}:
\STATE \hspace{0.5cm}Decode all {\em current load information} 
\STATE {\em Request Step}:
\STATE  \hspace{0.5cm}Compute  utility changes for all migrations 
\STATE  \hspace{0.5cm}Determine TP $\hat{b}$ corresponding to best change.
\STATE \hspace{0.5cm}Send association request  to TP $\hat{b}$ if relative improvement
 \STATE \hspace{0.5cm} is better than $\Delta$
\STATE  \hspace{0.5cm} \textbf{If} ACK received from TP $\hat{b}$
\STATE \hspace{0.8cm}Send request to release  to current TP
 \STATE \hspace{0.8cm}Send $w_k,R_{k,\hat{b}}(\rhob)$ to TP $\hat{b}$
\STATE  \hspace{0.5cm}\textbf{EndIf}
\STATE \textbf{Until}   ACK received from requested TP
\end{algorithmic}
\end{small}\vspace{-.8cm}
\end{table}

In  the appendix  we show that the proposed distributed LS stage provably converges and  the solution it yields upon convergence yields the same guarantees as in Proposition \ref{prop:ls2}. We note here that a distributed user-centric randomized algorithm has been recently proposed in \cite{bethan:arxiv}.  However, proving the convergence of that algorithm for arbitrary $\alpha$ remains an open problem.

%% file: abbreviations2bb.tex

\newcommand{\Reals}     {{{\mathrm I\!R}}}  
\newcommand{\Cplx}      {{{\mathsf I}\!\!\!{\mathrm C}}} 
\newcommand{\NCplx}     {{{\mathcal{CN}}}} 
\newcommand{\Ints}      {{{\mathbb Z}}} 
\newcommand{\Rats}      {{{\mathsf I}\!\!\!{\mathrm Q}}}    
\newcommand{\Nats}      {{{\mathrm{ I\!N}}}} 
\newcommand{\Tee}{      {{\mathbb T}}}
\newcommand{\ltwo}      {{\ell_2}}
\newcommand{\lone}      {{\ell_1}}
\newcommand{\Ltwo}      {{\mathbb L}^2}
\newcommand{\LtR}       {\Ltwo(\Reals)}
\newcommand{\Real}[1]   {{\mathrm{Re}\left\{#1\right\}}}
\newcommand{\Imag}[1]   {{\mathrm{Im}\left\{#1\right\}}}
\newcommand{\real}      {{\mathcal{R}} e}
\newcommand{\imag}      {{\mathcal{I}} m}

\newcommand{\ie}        {i.~e., \hspace{2pt}}      
\newcommand{\eg}        {e.~g., \hspace{2pt}}      %
\newcommand{\define}    {\stackrel{\scriptscriptstyle\triangle}{=}}  
\newcommand{\D}  {^{\dag}}               
\newcommand{\T}  {{\mathsf{T}}}          
\newcommand{\ct} {{\mathsf{*}}}          
\newcommand{\half}{{{^1\!\!\scriptscriptstyle{/}\!}_2}}       
\newcommand{\mrsss}[1]   {{\mathrm{\scriptscriptstyle{#1}}}}  
\newcommand{\sth}     {{\mathrm{th}}}    
\newcommand{\diag}    {{\mathrm{diag}}}  
\newcommand{\sgn}     {{\mathrm{sgn}}}   
\newcommand{\tr}      {{\mathrm{tr}}}    
\newcommand{\Res}     {{\mathrm{Res}}}   
\newcommand{\ld}      {{\mathrm{ld}}}    
\newcommand{\barg}    {\overline{\gamma}}     
\newcommand{\sqgb}    {\sqrt{\barg}}     
\newcommand{\e}       {{\mathrm e}}      
\newcommand{\E}       {{\mathrm E}}      
\newcommand{\lNZ}     {{\lambda_{\mathsf{NZ}}}} 
\newcommand{\Nt}      {{N_{\!\mrsss{T}}}}
\newcommand{\Nr}      {{N_{\!\mrsss{R}}}}
\newcommand{\uple}    {{\mathrm{ul}}}    
\newcommand{\upri}    {{\mathrm{ur}}}    
\newcommand{\lole}    {{\mathrm{ll}}}    
\newcommand{\lori}    {{\mathrm{lr}}}    

\newcommand{\Zrb}     {{\uwti 0}}      
\newcommand{\Oneb}    {{\uwti 1}}      

\newcommand{\uwti}[1]{{\mathbf #1}}
\newcommand{\ab}{{\uwti a}}  \newcommand{\Ab}{{\uwti A}}
\newcommand{\bb}{{\uwti b}}  \newcommand{\Bb}{{\uwti B}}
\newcommand{\cb}{{\uwti c}}  \newcommand{\Cb}{{\uwti C}}
\newcommand{\db}{{\uwti d}}  \newcommand{\Db}{{\uwti D}}
\newcommand{\eb}{{\uwti e}}  \newcommand{\Eb}{{\uwti E}}
\newcommand{\fb}{{\uwti f}}  \newcommand{\Fb}{{\uwti F}}
\newcommand{\gb}{{\uwti g}}  \newcommand{\Gb}{{\uwti G}}
\newcommand{\hb}{{\uwti h}}  \newcommand{\Hb}{{\uwti H}}
\newcommand{\ib}{{\uwti i}}  \newcommand{\Ib}{{\uwti I}}
\newcommand{\jb}{{\uwti j}}  \newcommand{\Jb}{{\uwti J}}
\newcommand{\kb}{{\uwti k}}  \newcommand{\Kb}{{\uwti K}}
\newcommand{\lb}{{\uwti l}}  \newcommand{\Lb}{{\uwti L}}
\newcommand{\mb}{{\uwti m}}  \newcommand{\Mb}{{\uwti M}}
\newcommand{\nb}{{\uwti n}}  \newcommand{\Nb}{{\uwti N}}
\newcommand{\ob}{{\uwti o}}  \newcommand{\Ob}{{\uwti O}}
\newcommand{\pb}{{\uwti p}}  \newcommand{\Pb}{{\uwti P}}
\newcommand{\qb}{{\uwti q}}  \newcommand{\Qb}{{\uwti Q}}
\newcommand{\rb}{{\uwti r}}  \newcommand{\Rb}{{\uwti R}}
\renewcommand{\sb}{{\uwti s}}\newcommand{\Sb}{{\uwti S}} 
\newcommand{\tb}{{\uwti t}}  \newcommand{\Tb}{{\uwti T}}
\newcommand{\ub}{{\uwti u}}  \newcommand{\Ub}{{\uwti U}}
\newcommand{\vb}{{\uwti v}}  \newcommand{\Vb}{{\uwti V}}
\newcommand{\wb}{{\uwti w}}  \newcommand{\Wb}{{\uwti W}}
\newcommand{\xb}{{\uwti x}}  \newcommand{\Xb}{{\uwti X}}
\newcommand{\yb}{{\uwti y}}  \newcommand{\Yb}{{\uwti Y}}
\newcommand{\zb}{{\uwti z}}  \newcommand{\Zb}{{\uwti Z}}

\newcommand{\alphab}      {{\bm \alpha}}          \newcommand{\Alphab}   {{\uwti {\mathrm A}}}
\newcommand{\betab}       {{\bm \beta}}           \newcommand{\Betab}    {{\uwti {\mathrm B}}}
\newcommand{\gammab}      {{\bm \gamma}}          \newcommand{\Gammab}   {{\bm \Gamma}}
\newcommand{\deltab}      {{\bm \delta}}          \newcommand{\Deltab}   {{\bm \Delta}}
\newcommand{\epsilonb}    {{\bm \epsilon}}        \newcommand{\Epsilonb} {{\uwti {\mathrm E}}}
\newcommand{\varepsilonb} {{\bm \varepsilon}}
\newcommand{\vepsilonb}   {{\bm \varepsilon}}
\newcommand{\zetab}       {{\bm \zeta}}           \newcommand{\Zetab}    {{\uwti {\mathrm Z}}}
\newcommand{\etab}        {{\bm \eta}}            \newcommand{\Etab}     {{\uwti {\mathrm H}}}
\newcommand{\thetab}      {{\bm \theta}}          \newcommand{\Thetab}   {{\bm \Theta}}
\newcommand{\varthetab}   {{\bm \vartheta}}
\newcommand{\vthetab}     {{\bm \vartheta}}
\newcommand{\iotab}       {{\bm \iota}}           \newcommand{\Iotab}    {{\uwti {\mathrm I}}}
\newcommand{\kappab}      {{\bm \kappa}}          \newcommand{\Kappab}   {{\uwti {\mathrm K}}}
\newcommand{\lambdab}     {{\bm \lambda}}         \newcommand{\Lambdab}  {{\bm \Lambda}}
\newcommand{\mub}         {{\bm \mu}}             \newcommand{\Mub}      {{\uwti {\mathrm M}}}
\newcommand{\nub}         {{\bm \nu}}             \newcommand{\Nub}      {{\uwti {\mathrm N}}}
\newcommand{\xib}         {{\bm \xi}}             \newcommand{\Xib}      {{\bm \Xi}}
\newcommand{\omicronb}    {{\uwti {\mathrm o}}} \newcommand{\Omicronb} {{\uwti {\mathrm O}}}
\newcommand{\pib}         {{\bm \pi}}             \newcommand{\Pib}      {{\bm \Pi}}
\newcommand{\varpib}      {{\bm \varpi}}
\newcommand{\vpib}        {{\bm \varpi}}
\newcommand{\rhob}        {{\bm \rho}}            \newcommand{\Rhob}     {{\uwti {\mathrm P}}}
\newcommand{\varrhob}     {{\bm \varrho}}
\newcommand{\vrhob}       {{\bm \varrho}}
\newcommand{\sigmab}      {{\bm \sigma}}          \newcommand{\Sigmab}   {\uwti{\mathnormal\Sigma}}
\newcommand{\varsigmab}   {{\bm \varsigma}}
\newcommand{\vsigmab}     {{\bm \varsigma}}
\newcommand{\taub}        {{\bm \tau}}            \newcommand{\Taub}     {{\uwti {\mathrm T}}}
\newcommand{\upsilonb}    {{\bm \upsilon}}        \newcommand{\Upsilonb} {{\bm \Upsilon}}
\newcommand{\phib}        {{\bm \phi}}            \newcommand{\Phib}     {{\bm \Phi}}
\newcommand{\varphib}     {{\bm \varphi}}
\newcommand{\vphib}       {{\bm \varphi}}
\newcommand{\chib}        {{\bm \chi}}            \newcommand{\Chib}     {{\uwti {\mathrm X}}}
\newcommand{\psib}        {{\bm \psi}}            \newcommand{\Psib}     {{\bm \Psi}}
\newcommand{\omegab}      {{\bm \omega}}          \newcommand{\Omegab}   {{\bm \Omega}}

\newcommand{\Ac} {{\mathcal A}}         \newcommand{\Ak} {{\bm {\mathcal A}}}
\newcommand{\Bc} {{\mathcal B}}         \newcommand{\Bk} {{\bm {\mathcal B}}}
\newcommand{\Cc} {{\mathcal C}}         \newcommand{\Ck} {{\bm {\mathcal C}}}
\newcommand{\Dc} {{\mathcal D}}         \newcommand{\Dk} {{\bm {\mathcal D}}}
\newcommand{\Ec} {{\mathcal E}}         \newcommand{\Ek} {{\bm {\mathcal E}}}
\newcommand{\Fc} {{\mathcal F}}         \newcommand{\Fk} {{\bm {\mathcal F}}}
\newcommand{\Gc} {{\mathcal G}}         \newcommand{\Gk} {{\bm {\mathcal G}}}
\newcommand{\Hc} {\uwti{\mathcal H}}         \newcommand{\Hk} {{\bm {\mathcal H}}}
\newcommand{\Ic} {{\mathcal I}}         \newcommand{\Ik} {{\bm {\mathcal I}}}
\newcommand{\Jc} {{\mathcal J}}         \newcommand{\Jk} {{\bm {\mathcal J}}}
\newcommand{\Kc} {{\mathcal K}}         \newcommand{\Kk} {{\bm {\mathcal K}}}
\newcommand{\Lc} {{\mathcal L}}         \newcommand{\Lk} {{\bm {\mathcal L}}}
\newcommand{\Mc} {{\mathcal M}}         \newcommand{\Mk} {{\bm {\mathcal M}}}
\newcommand{\Nc} {{\mathcal N}}         \newcommand{\Nk} {{\bm {\mathcal N}}}
\newcommand{\Oc} {{\mathcal O}}         \newcommand{\Ok} {{\bm {\mathcal O}}}
\newcommand{\Pc} {{\mathcal P}}         \newcommand{\Pk} {{\bm {\mathcal P}}}
\newcommand{\Qc} {{\mathcal Q}}         \newcommand{\Qk} {{\bm {\mathcal Q}}}
\newcommand{\Rc} {{\mathcal R}}         \newcommand{\Rk} {{\bm {\mathcal R}}}
\newcommand{\Sc} {{\mathcal S}}         \newcommand{\Sk} {{\bm {\mathcal S}}}
\newcommand{\Tc} {{\mathcal T}}         \newcommand{\Tk} {{\bm {\mathcal T}}}
\newcommand{\Uc} {{\mathcal U}}         \newcommand{\Uk} {{\bm {\mathcal U}}}
\newcommand{\Vc} {{\mathcal V}}         \newcommand{\Vk} {{\bm {\mathcal V}}}
\newcommand{\Wc} {{\mathcal W}}         \newcommand{\Wk} {{\bm {\mathcal W}}}
\newcommand{\Xc} {{\mathcal X}}         \newcommand{\Xk} {{\bm {\mathcal X}}}
\newcommand{\Yc} {{\mathcal Y}}         \newcommand{\Yk} {{\bm {\mathcal Y}}}
\newcommand{\Zc} {{\mathcal Z}}         \newcommand{\Zk} {{\bm {\mathcal Z}}}

\newcommand{\Grb}   {{\uwti {\mathrm V}}}
\newcommand{\Pulk} {{\underline{{\bm {\mathcal P}}}}}
\newcommand{\Qulk} {{\underline{{\bm {\mathcal Q}}}}}
\newcommand{\Culk} {{\underline{{\bm {\mathcal C}}}}}
\newcommand{\Mulk} {{\underline{{\bm {\mathcal M}}}}}
\newcommand{\Bulk} {{\underline{{\bm {\mathcal B}}}}}
\newcommand{\Tulk} {{\underline{{\bm {\mathcal T}}}}}
\newcommand{\Iulk} {{\underline{{\bm {\mathcal I}}}}}
\newcommand{\Fulk} {{\underline{{\bm {\mathcal F}}}}}
\newcommand{\Aulk} {{\underline{{\bm {\mathcal A}}}}}
\newcommand{\Wulk} {{\underline{{\bm {\mathcal W}}}}}

\newcommand{\Aulc} {{\underline{\mathcal A}}}
\newcommand{\Bulc} {{\underline{\mathcal B}}}
\newcommand{\Iulc} {{\underline{\mathcal I}}}
\newcommand{\Julc} {{\underline{\mathcal J}}}
\newcommand{\Gulc} {{\underline{\mathcal G}}}
\newcommand{\Culc} {{\underline{\mathcal C}}}
\newcommand{\Pulc} {{\underline{\mathcal P}}}
\newcommand{\Lulc} {{\underline{\mathcal L}}}
\newcommand{\Uulc} {{\underline{\mathcal U}}}
\newcommand{\Eulc} {{\underline{\mathcal E}}}
\newcommand{\Fulc} {{\underline{\mathcal F}}}
\newcommand{\Rulc} {{\underline{\mathcal R}}}
\newcommand{\Sulc} {{\underline{\mathcal S}}}
\newcommand{\Aul}  {{\underline A}}              \newcommand{\aul}  {{\underline a}}
\newcommand{\Bul}  {{\underline B}}              \newcommand{\bul}  {{\underline b}}
\newcommand{\Cul}  {{\underline C}}              \newcommand{\cul}  {{\underline c}}
\newcommand{\Dul}  {{\underline D}}              \newcommand{\dul}  {{\underline d}}
\newcommand{\Eul}  {{\underline E}}              \newcommand{\eul}  {{\underline e}}
\newcommand{\Ful}  {{\underline F}}              \newcommand{\ful}  {{\underline f}}
\newcommand{\Gul}  {{\underline G}}              \newcommand{\gul}  {{\underline g}}
\newcommand{\Hul}  {{\underline H}}              \newcommand{\hul}  {{\underline h}}
\newcommand{\Iul}  {{\underline I}}              \newcommand{\iul}  {{\underline i}}
\newcommand{\Jul}  {{\underline J}}              \newcommand{\jul}  {{\underline j}}
\newcommand{\Kul}  {{\underline K}}              \newcommand{\kul}  {{\underline k}}
\newcommand{\Lul}  {{\underline L}}              \newcommand{\lul}  {{\underline l}}
\newcommand{\Mul}  {{\underline M}}              \newcommand{\mull} {{\underline m}}
\newcommand{\Nul}  {{\underline N}}              \newcommand{\nul}  {{\underline n}}
\newcommand{\Oul}  {{\underline 0}}              \newcommand{\oul}  {{\underline o}}
\newcommand{\Pul}  {{\underline P}}              \newcommand{\pul}  {{\underline p}}
\newcommand{\Qul}  {{\underline Q}}              \newcommand{\qul}  {{\underline q}}
\newcommand{\Rul}  {{\underline R}}              \newcommand{\rul}  {{\underline r}}
\newcommand{\Sul}  {{\underline S}}              \newcommand{\sul}  {{\underline s}}
\newcommand{\Tul}  {{\underline T}}              \newcommand{\tul}  {{\underline t}}
\newcommand{\Uul}  {{\underline U}}              \newcommand{\uul}  {{\underline u}}
\newcommand{\Vul}  {{\underline V}}              \newcommand{\vul}  {{\underline v}}
\newcommand{\Wul}  {{\underline W}}              \newcommand{\wul}  {{\underline w}}
\newcommand{\Xul}  {{\underline X}}              \newcommand{\xul}  {{\underline x}}
\newcommand{\Yul}  {{\underline Y}}              \newcommand{\yul}  {{\underline y}}
\newcommand{\Zul}  {{\underline Z}}              \newcommand{\zul}  {{\underline z}}
\newcommand{\Omegaul}  {{\underline \Omega}}
\newcommand{\mc}[1]     {{\mathcal{#1}}}        
\newcommand{\what}[1]   {\widehat{#1}}          
\newcommand{\wtld}[1]   {\widetilde{#1}}        
\newcommand{\Portho}[1] {\Pb_{#1}^{\perp}}      
\newcommand{\ba}{\begin{array}}
\newcommand{\ea}{\end{array}}

%% file: GPwnE.tex
	\section{AF optimization}
	\label{AFO}

	The association scheme described in the previous section determines  $\Uc^{(b)}$, the set of users associated  to TP $b$ for all $b\in\Bc$.
In this section, for a given user association, we present a centralized algorithm to
determine $\rho_{b}$ for each $b$ so as to optimize the   system utility over different $\alpha$ regimes. 
For brevity we suppose that $\alpha>1$. The analogous results   for all other $\alpha$ values  as well as an equivalent distributed variant of the proposed approach  are deferred to the appendix.
      The     AF optimization problem in this  regime is given by
		\begin{eqnarray}
	\mbox{min}_{{\boldsymbol{\rho}}\in{[0,1]^B} } & \left\{\sum_{b\in{B}}{\left(\sum_{k\in{\Uc^{(b)}}}{ \tilde{w}_k/(R_{k,b}(\rhob))^{1-1/\alpha}}\right)^\alpha}\right\}
           \label{origbetagreater1}
	\end{eqnarray}
	where $\tilde{w}_k= (\frac{w_k}{\alpha-1})^{1/\alpha}$ and  $R_{k,b}(\rhob)$ is given by (\ref{eq:avrateI2}).
 We let $\betab_k=\{\beta_{k,b}\}\;\forall{b}\in{B}$ denote the vector containing all fading coefficients pertaining to user $k$ on any slot. 
Then, we introduce auxiliary
 variables $g_{k,b}(\betab_k)$ for each vector $\betab_k$ for each user $k\in\Uc^{(b)}$ for each TP $b$.
Using $g_{k,b}(\betab_k)$ as a filter at user $k$ to detect the  signal transmitted from TP $b$  over that slot, the
 mean squared   error (MSE),  $e_{k,b}(\betab_k,\rhob)$, is given by
	\begin{eqnarray}
\begin{split}
	&e_{k,b}(\betab_k,\rhob)=\left|g_{k,b}(\betab_k)\sqrt{\beta_{k,b}}-1\right|^2+\left|g_{k,b}(\betab_k)\right|^2\\&\;\;\;\;\;\;\;\;\;\;\;\;\;+\left|g_{k,b}(\betab_k)\right|^2 \sum_{b'\neq b}{\beta_{k,b'}\rho_{b'}}
\end{split}
           \label{error}
	\end{eqnarray}
	Using the   mutual information and MSE identity and introducing more auxiliary variables (cf. \cite{rajiv_agarwal}),  we have 
	\begin{eqnarray}
\nonumber 	R_{k,b}(\rhob)=\rho_b\mathbb{E} [\mbox{max}_{g_{k,b}(\betab_k),s_{k,b}(\betab_k)\geq {0}}   \\\;\;\;\;\;\{1-s_{k,b}(\betab_k)e_{k,b}(\betab_k,\rhob)+\log(s_{k,b}(\betab_k)) \}]
          \label{altrate}
	\end{eqnarray}
	The solution of each inner maximization problem in (\ref{altrate}) is obtained by setting $g_{k,b}(\betab_k)$ to be the MMSE filter $\hat{g}_{k,b}(\betab_k)$ with $s_{k,b}(\betab_k)=\hat{s}_{k,b}(\betab_k)=1/\hat{e}_{k,b}(\betab_k,\rhob)$, where $\hat{e}_{k,b}(\betab_k,\rhob)=e_{k,b}(\betab_k,\rhob)\mid_{g_{k,b}(\betab_k)=\hat{g}_{k,b}(\betab_k)}$.
	Using (\ref{altrate}), the   problem in (\ref{origbetagreater1}) (for the given association) can be re-formulated as the following optimization problem over variables $\boldsymbol{\rho}$,
$\boldsymbol{s}=\{s_{k,b}(\betab_k)\},\boldsymbol{g}=\{g_{k,b}(\betab_k)\}\;\forall \betab_k,k\in{\Uc^{(b)}},{b}\in{B}$.
\begin{eqnarray}\label{genbetagreater1}
\min_{{\boldsymbol{\rho}}\in{\bf{[0,1]}},{\bf g}\geq{\bf 0},{\bf s}\geq{\bf 1} } \left\{
 \sum_{b\in{B}}\left(\sum_{k\in{\Uc^{(b)}}}\;\;\;\;\;\;\;\;\;\;\;\;\;\;\;\;\;\;\;\;\;\;\;\;\;\;\;\;\;\;\;\;\;\;\;\;\;\;\;\;\;\; \right.\right.\\
 \nonumber \left.\left. \frac{\tilde{w}_k}{(\rho_b\mathbb{E}[1-s_{k,b}(\betab_k)e_{k,b}(\betab_k,\rhob)+\log(s_{k,b}(\betab_k))])^{1-1/\alpha}}\right)^\alpha\right\}
	\end{eqnarray}
	Note that for a fixed $\rhob$, (\ref{genbetagreater1}) can be optimized over $\bf{s,g}$  via the closed form expressions given above. On the other hand, for fixed $\bf{s,g}$ to optimize (\ref{genbetagreater1}) over $\rhob$, we introduce additional variables $\boldsymbol{z}=\{z_{b}\}\;\forall{b}\in{B}$ and $\boldsymbol{t}=\{t_{k,b}\},\;\forall\; k\in{\Uc^{(b)}},b\in{B}$ and  express the reduced problem in (\ref{genbetagreater1}) as
	\begin{equation}
\begin{split}
	&\mbox{min}_{{\boldsymbol{\rho}}\in{\bf{[0,1]}},{\bf z}\geq{\bf 0},{\bf t}\geq{\bf 0} }\left\{\sum_{b\in{B}}{z_{b}^\alpha}\right\}\\
	&\mbox{subject to}  \\
	 &z_{b}\geq{\sum_{k\in{\Uc^{(b)}}}{ \tilde{w}_kt_{k,b}^{1/\alpha-1}}}\ \ \forall{k,b}\\
	 & t_{k,b}\leq{\rho_{b}\mathbb{E}[1-s_{k,b}(\betab_k)e_{k,b}(\betab_k,\rhob)+\log(s_{k,b}(\betab_k))]}\ \ \forall{k,b}
\label{eqgenbetagreater1}
\end{split}
	\end{equation}
		Notice that  (\ref{eqgenbetagreater1}) can in turn be re-written as
	\begin{equation}
\begin{split}
	&\mbox{min}_{{\boldsymbol{\rho}}\in{\bf{[0,1]}},{\bf z}\geq{\bf 0},{\bf t}\geq{\bf 0} }\{\sum_{b\in{B}}{z_{b}^\alpha}\}\\
	&\mbox{subject to}\\
	 &\sum_{k}{z_{b}^{-1} \tilde{w}_kt_{k,b}^{1/\alpha-1}}\leq{1}\ \ \forall{k,b}\\
	 &\frac{ t_{k,b}\rho_{b}^{-1}+\mathbb{E}[s_{k,b}(\betab_k)e_{k,b}(\betab_k,\rhob)]}{1+\mathbb{E}[\log(s_{k,b}(\betab_k))]}\leq1\ \ \forall{k,b}
\label{fixbetagreater1}
\end{split}
	\end{equation}
	The problem in (\ref{fixbetagreater1}) is a geometric program (GP) since all constraints are inequalities involving posynomials.
	                            Thus, we can  repeat the following two steps
until  convergence.
\begin{enumerate}
         \item  Fix $\boldsymbol{\rho}$ and minimize (\ref{genbetagreater1}) over $\boldsymbol{s,g}$ using closed form solution of (\ref{altrate}).
	\item Fix $\boldsymbol{s,g}$ and minimize (\ref{genbetagreater1}) over $\boldsymbol{\rho}$ by solving equivalent GP in (\ref{fixbetagreater1}).
\end{enumerate}
Note that in the described auxiliary function method we have a monotonic improvement in the objective value of (\ref{genbetagreater1}) so that convergence is guaranteed. 

%% file: jointlbasscv2wn.tex
\section{Joint Association \& AF optimization}
\label{JAAFO}

\begin{figure}[h]
\centering
\includegraphics[width=0.4\textwidth]{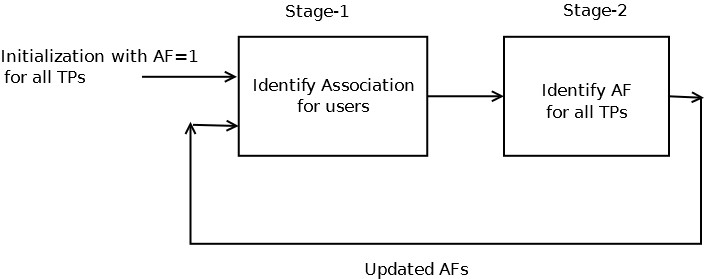}
\caption{Joint Association and AF optimization block diagram}
\label{block1}
\end{figure}
We propose two joint association \& AF optimization algorithms for solving the    problem in (\ref{eq:BJigoriginal}). These algorithms   follow an alternating optimization approach where user association (stage-1) and AF (stage-2) are optimized  in an alternating fashion. Fig.~\ref{block1} shows a block-level decomposition.  
The first algorithm is the Joint GLS-AF algorithm, in which we first run the GLS algorithm (Algorithm in Table \ref{algo:lb}) and use the obtained association in our AF optimization algorithm in  Section IV .
We repeat the following two steps   until the benefit in terms of the alpha-fairness system utility falls below a threshold. 
\begin{enumerate}
         \item{Stage1}--Fix $\boldsymbol{\rho}$ and use GLS algorithm to calculate the user association.
	\item{Stage2}--Fix the association  and optimize over $\boldsymbol{\rho}$ using the auxiliary function method given in  Section IV .
\end{enumerate}
It is evident that both stages in the above alternating approach can be performed using the respective distributed versions that we derived before. However, one issue with the proposed joint GLS-AF algorithm, is that the TPs that do not serve any user in any one iteration will be discarded in all subsequent iterations. To overcome this potential limitation, we consider the joint relaxed association and AF (Joint RA-AF) algorithm. To obtain the association, this latter algorithm in   stage-1 solves the convex optimization problem obtained by relaxing variables $x_{k,b},\;\forall\;k,b$ in (\ref{eq:bg1}) or (\ref{eq:bg2}) to be continuous variables in $[0,1]$. In this solution, a user $k$ can have $x_{k,b}$ non-zero for more than one TP $b$. In stage-2, the algorithm fixes $x_{k,b}$ for all $k,b$ and optimizes the AF. To do so, it uses the auxiliary function method of   Section IV  on the objective in the problem (\ref{eq:bg1}) rather than (\ref{origbetagreater1}) as $x_{k,b}$ can now have fractional values.
This two stage procedure is repeated until the  benefit in system utility falls below a threshold. Finally, the Joint RA-AF algorithm rounds $x_{k,b}$ to obtain a feasible association.

%% file: Evaluationv2wn.tex
\section{Evaluation}

We present a detailed evaluation of our proposed: Greedy Local Search (GLS) algorithm, the distributed  Greedy (DG) algorithm and the joint association \& AF optimization  algorithms over an LTE HetNet deployment. In our evaluation topology  an enhanced NodeB (eNB) covers  the coordination area. The eNB site comprises of three cells (sectors), where each sector contains a set of eleven TPs  formed by one macro   and ten lower power (pico) nodes. We drop ninety nine users 
 on the eNB site so there are a total of $B=33$ TPs and $K=99$ users. All TPs  and users have a single antenna each. 
  We employ the conservative rates and ignore fast fading in the  results presented  in  Section VI-A \&  Section VI-B.
The results  incorporating actual rates, fast fading and efficient per-slot user scheduling are presented later in Section VI-C. 

%
\subsection{Association}\label{Association}
We compare the  GLS \& DG algorithms proposed in Section III-A  and Section III-B, respectively, to the following:
\setlength\itemindent{0pt}
\begin{itemize}[leftmargin=*]
  \setlength{\parskip}{0pt}
  \setlength{\itemsep}{0pt plus 1pt}
\item{Relaxed Upperbound (RU)}--Solves the convex optimization problem obtained by relaxing $x_{k,b}$ in (\ref{eq:bg1}) or (\ref{eq:bg2}). 
  Though the obtained solution need not be feasible for (\ref{eq:Bigoriginal}), the scheme provides us with an upperbound  on the optimal of (\ref{eq:Bigoriginal}).
\item Relaxed Rounded Association (RRA)--Solves the convex optimization problem obtained by relaxing $x_{k,b}$ in (\ref{eq:bg1}) or (\ref{eq:bg2}). Each user $k$
connects to the TP $b$ corresponding to highest $x_{k,b}$ in the obtained convex optimization solution. This scheme is widely used to represent the performance that can be achieved by a feasible and near-optimal user association scheme. However, it requires solving
a convex problem that can be   computationally quite complex   compared to   GLS   in a dense deployment.
\item Max SNR Association (MSA)-- Each user independently connects to the TP from which it sees the highest average channel gain. This scheme is the most common baseline. 
\end{itemize}
We evaluate the association algorithms by examining their returned
utility function values for varying $\alpha$. We also
evaluate the additional gain yielded by the local search (LS) stage over the greedy one in the GLS algorithm.

\subsubsection{$\alpha\leq1$}
We begin with an evaluation of   GLS   and the distributed greedy (DG) algorithm  in the regime $\alpha\leq1$, where we consider the maximization of the objective in (\ref{eq:bg2}).
We set $\rho=1$ for each of the 33 TPs and list the   utility values of different association
 algorithms in Table IV. As suggested by the guarantee in Proposition \ref{prop:bd11}, we observe that greedy  stage of   GLS   itself  performs very close to the upper bound RU, and hence close to the optimal and provides good gains over the MSA scheme. Notice that   GLS     outperforms the RRA despite having a much lower computational complexity.   Moreover, the DG algorithm performs close to the former two ones, while simultaneously offering the benefits of a distributed implementation. We also observe   that the local search iterations (LSIs) of   GLS     are at-most 1  
 and that there is a slight utility gain obtained by the LS stage.  
Interestingly, upon employing the association algorithm  from \cite{prasad:joint} 
we observed that the GLS indeed yields the optimal association for this example when $\alpha=1$. 

\begin{table}
\begin{center}
\scalebox{0.7}{
\begin{tabular}{c|c|c|c|c|c|c|c}

$\alpha$ & Greedy & GLS & RU & RRA & MSA & DG & LSI\\
  \hline
0.25 & 67.75 & 67.82 & 67.82 & 67.82 & 65.08 &67.48 & 1\\
0.5 & 112.67 & 112.67 & 112.71 & 112.52 & 107.03 &110.39 & 0\\
0.75 & 288.57 & 288.57 & 288.82 & 288.46 & 277.65&283.98 & 0\\
1.0 & -133.93 & -133.87 & -133.3 1& -133.93 & -154.67 &-139.76& 1\\
\end{tabular}
}
\caption{\label{perfbetaless=1}Utility versus $\alpha$}

\end{center}
\end{table}

\subsubsection{$\alpha>1$}
 Next we study the performance of GLS \& DG algorithms in $\alpha>1$ region, where we consider the minimization of the objective in (\ref{eq:bg1}). As seen in Fig. 2(a) the proposed GLS \& DG   perform very similarly and they noticeably outperform  RRA in $\alpha>3$ regime while beating MSA over the entire range of $\alpha>1$. For example, GLS performs 13.5 \% better than RRA and 80\% better than MSA at $\alpha=4$. MSA performs poorly throughout the $\alpha>1$ regime since it has a naive user specific view rather than an optimized system specific  view. The superiority of GLS \& DG over RRA \& MSA increases with increase in $\alpha$. For example, at a high $\alpha=10$, which approaches max-min fairness, the GLS outperforms RRA \& MSA by 93.2\% and 100\% respectively. \\
 In Table V we study the advantage of doing local search in the $\alpha>1$ region. It is known that the greedy algorithm does not yield  a constant factor approximation for the constrained minimization of a non-negative non-decreasing supermodular set function. \footnote{This problem is equivalent to the constrained maximization of a   submodular set function albeit where that set function is not non-negative and non-decreasing, so that the classical result \cite{nemhaus:algo} is inapplicable.}
Therefore, the greedy stage need not be close to the optimal and there is room for improvement by the
LS stage. As seen in Table V, though the number of LS iterations are at-most 2, the order of gain over the greedy is upto 3.6\%. At a higher $\alpha=10$ the gain of GLS over greedy  shoots up to 43\%, with the number of LS iterations  equal to 5. Therefore, as $\alpha$ is progressively increased, the  local search stage of the GLS algorithm becomes increasingly important.

\begin{table}

\begin{center}
\scalebox{0.7}{
\begin{tabular}{c|c|c|c}
$\alpha$ & Greedy & GLS &  LSI\\
  \hline
1.25 &563.9 &563.9  &0 \\
1.5 & 411.4 &411.3  &1\\
1.75 & 408.7 &406.8  &2\\
2.0 &  462.6 &458.9  &2\\
2.25 & 565.6  &559.0 &2\\
2.5 & 728.5 &717.2  &2\\
\end{tabular}
\begin{tabular}{c|c|c|c}
$\alpha$ & Greedy & GLS &  LSI\\
  \hline
2.75 & 975.2 &956.1  &2\\
3.0 & 1345.8 &1314.2  &2\\
3.25 & 1904.6 &1853.0  &2\\
3.5 & 2754.6 &2671.2  &2\\
3.75 & 4045.1 &3911.4  &2\\
4.0 & 5953.6 &5740.7  &2\\
\end{tabular}
}
\caption{\label{perfbetagreater1}Local Search Improvement}\vspace{-.5cm}
\end{center}
\end{table}


\begin{figure}
  \begin{subfigure}[b]{.45\linewidth}
    \includegraphics[width=1.1\linewidth]{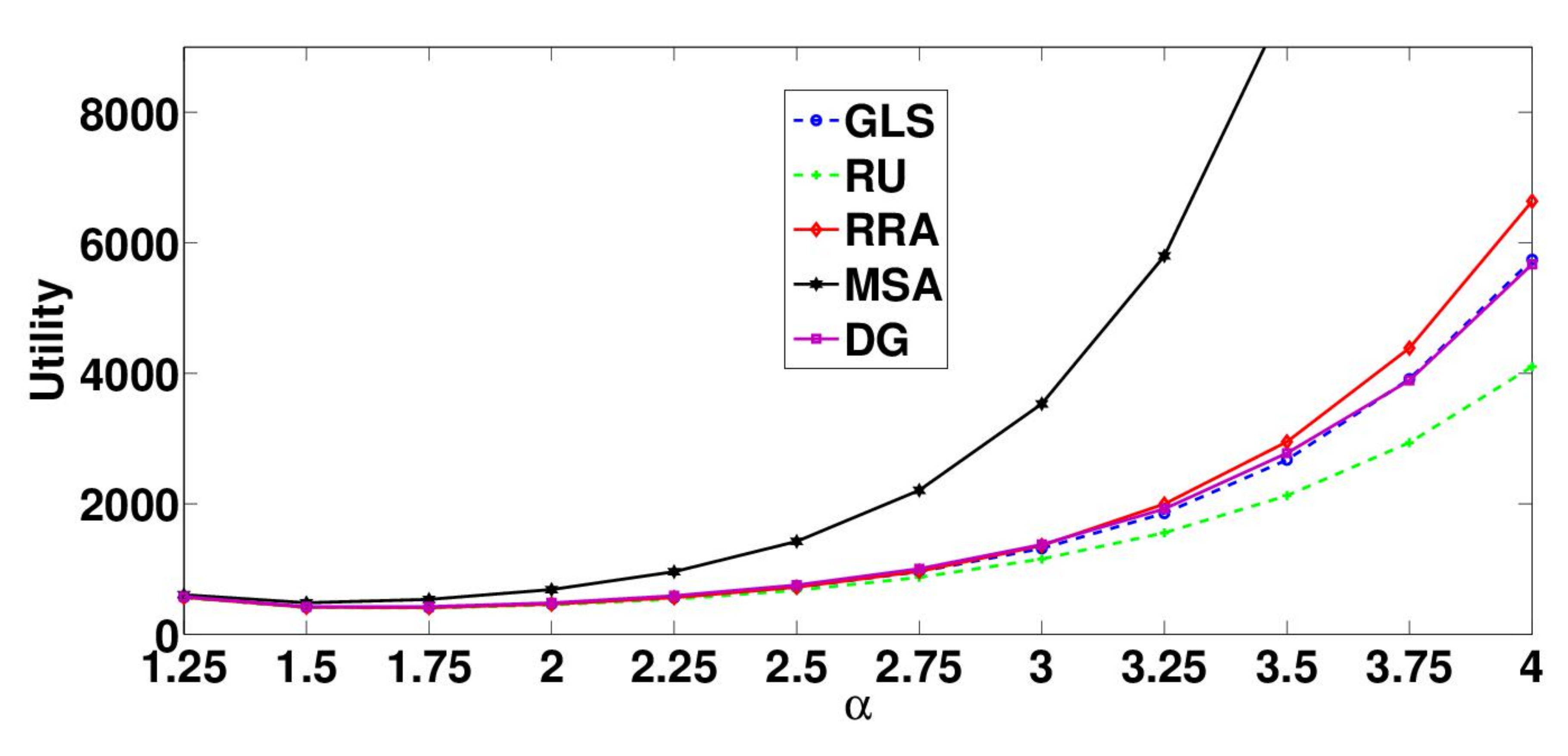}
    \caption*{2(a) Utility vs $\alpha$}
    \label{exp1}
  \end{subfigure}\hspace{.3cm}
  \begin{subfigure}[b]{.45\linewidth}
    \includegraphics[width=1.1\linewidth]{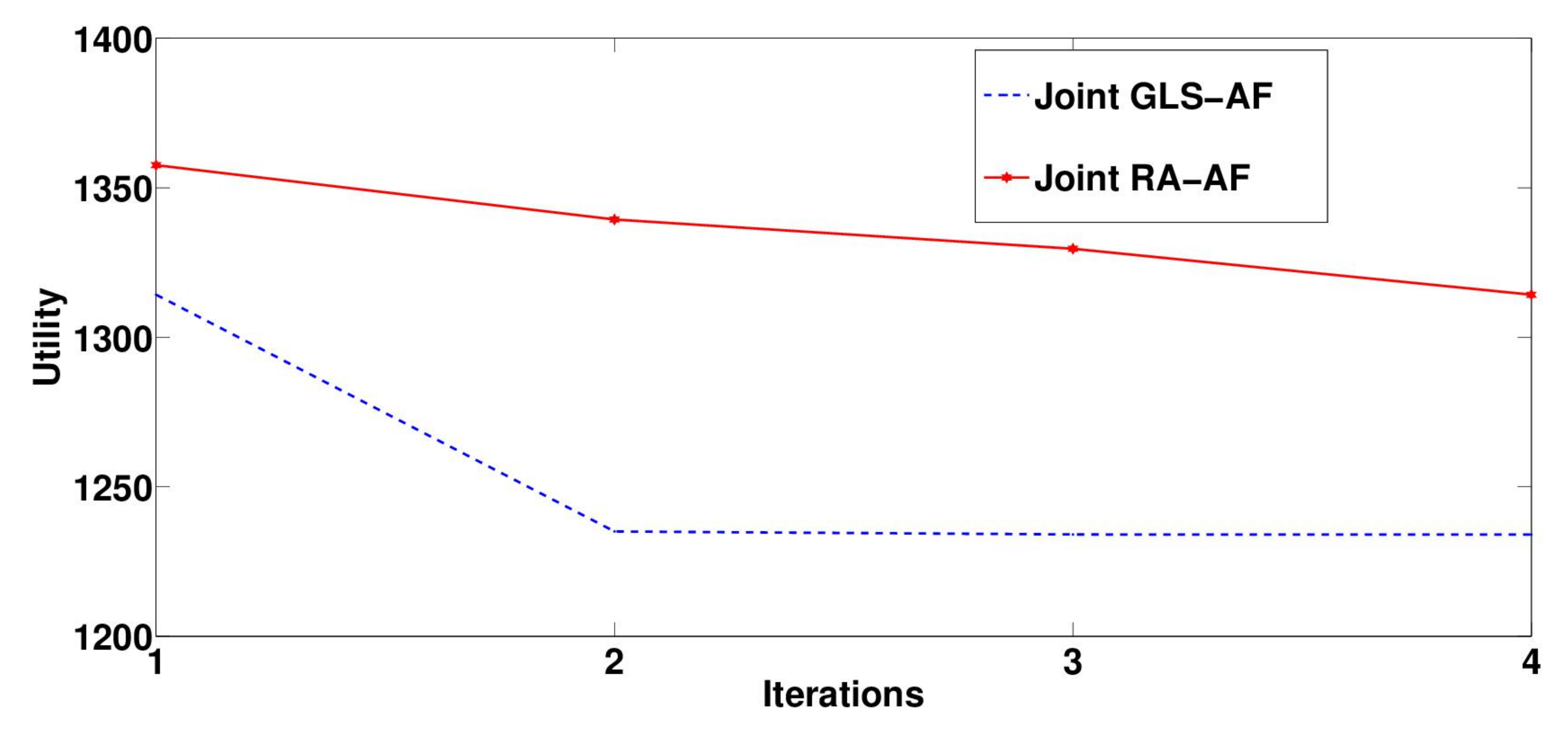}
    \caption*{2(b) Utility vs iterations}
    \label{JALB}
  \end{subfigure}\vspace{-.4cm}
\end{figure}

\subsection{Joint Association \& Activation fraction optimization}
\label{JAAF}
%
%

In  Fig. 2(b) we study the performance of the two joint algorithms described in Section V  for $\alpha=3.0$ for up-to 4 iterations. Each point in the plot corresponds to an iteration, and is the  utility value obtained using the updated association, where that association itself is calculated using the updated value of the activation fractions. The value at the first iteration is the utility corresponding to the association done using AF equal to 1 for all TPs. In the Joint RA-AF, at every iteration we calculate the utility by rounding the fractional association as done in the RRA algorithm. However, as mentioned in Section V, fractional values of the association variables $\{x_{k,b}\}$ are passed on to its second stage of AF identification. MSA with $\rho=1$ for each TP with a utility value of 3531.8, performs much worse than the Joint GLS-AF \& Joint RA-AF schemes. We obtain a gain of 6.1\% for Joint GLS-AF over the case when we do only  association via GLS with a fixed $\boldsymbol{\rho=1}$, which demonstrates the benefit of doing the joint association and AF optimization. The Joint RA-AF scheme performs worse (upto 8.45\%) than the Joint GLS-AF algorithm at every iteration, illustrating that the benefits of GLS over RRA observed before at $\boldsymbol{\rho=1}$ are preserved even in the joint optimization problem. \\ For $\alpha=0.5$, Joint GLS-AF  performs 23.36\% better than MSA with $\boldsymbol{\rho=1}$,  as compared to the gain of 4.6\% obtained by GLS over MSA observed in Table IV, again demonstrating the gain of optimizing AF and the association jointly. 
 We observe that Joint GLS-AF \& Joint RRA-AF algorithms perform very close to each other in $\alpha<1$ regime. This is because of the similar  performance of GLS and RRA schemes in this $\alpha$ regime.
\subsection{ Result Verification with Fast Fading}\label{ResultVerification}
Finally, in this section we incorporate fast fading and efficient per-slot user scheduling  to asses the benefits   of the association and activation fractions  calculated using proposed Joint GLS-AF algorithm.
In particular, we assume that each frame comprises of $5000$ slots and model all fast fading coefficients seen by each user on each slot as i.i.d. complex normal $\Cc\Nc(0,1)$ variables. We   randomly generate an ON-OFF pattern (for slots across each frame)  for each TP that is compliant with its assigned activation fraction. Further, each TP employs the per-slot gradient based  scheduling policy \cite{stol:grad} over the set of users associated to it in order to maximize the utility. Then, using the  actual per-user average rates so obtained, we compute the system utility values for different schemes. For $\alpha=0.5$
we observed that the Joint GLS-AF scheme yields a $15.35 \%$ gain over the baseline scheme (MSA with $\rhob=\Oneb$), while the gain of the GLS with $\rhob=\Oneb$ over the baseline is $5.32\%$.
 For $\alpha=3$ the gains of these two schemes over the baseline are $47.8\%$ and $39.4\%$, respectively.
This validates that our   approach to obtain the association and AF does indeed result in significant gains in the presence of fast fading and efficient fine time-scale (per-slot) scheduling.

\section{Conclusion}
We   analyzed and evaluated novel
association and activation fraction optimization algorithms for
maximizing the alpha-fairness utility in HetNets.
We derived useful performance guarantees and demonstrated the significant benefits of our  proposed algorithms over a practical HetNet topology.


%% file: ExtendedsubmissionWiOpt15.bbl
\begin{thebibliography}{10}

\bibitem{standard:3gpp_TR36872}
{3GPP}, ``Study on small cell enhancements for {E-UTRA} and {E-UTRAN} –
  physical-layer aspects,'' {\em TR36.872 V12.0.0}, Sept. 2013.

\bibitem{gesbert:binary}
A.~Gjendemsjoe, D.~Gesbert, G.~Oien, and S.~Kiani, ``Binary power control for
  sum rate maximization over multiple interfering links,'' {\em IEEE Trans.
  Wireless. Comm.}, Aug. 2008.

\bibitem{Wei:INFO2011}
W.~Yu, T.~Kwon, and C.~Shin, ``Multicell coordination via joint scheduling,
  beamforming and power spectrum adaptation,'' in {\em Proc. IEEE INFOCOM},
  pp.~2570--2578, Apr. 2011.

\bibitem{Lozano:IA}
O.~Ayach~El, A.~Lozano, and R.~Heath, ``On the overhead of interference
  alignment: Training, feedback, and cooperation,'' {\em IEEE Trans. on
  Wireless Comm.}, Nov. 2012.

\bibitem{Huang:MBC}
Y.~Huang, G.~Zheng, M.~Bengtsson, K.-K. Wong, L.~Yang, and B.~Ottersten,
  ``Distributed multicell beamforming with limited intercell coordination,''
  {\em IEEE Trans. on Sig. Proc.}, Jan. 2011.

\bibitem{Sanjabi:joint}
M.~Sanjabi, M.~Razaviyayn, and Z.~Q. Luo, ``Optimal joint base station
  assignment and beamforming for heterogeneous networks,'' {\em IEEE Trans. on
  Sig. Proc.}, Apr. 2014.

\bibitem{tajer:robust}
A.~Tajer, N.~Prasad, and X.~Wang, ``Robust linear precoder design for
  multi-cell downlink transmission,'' {\em IEEE Trans. on Signal Processing},
  jan 2011.

\bibitem{koshy:twoTS}
N.~Vaidhiyan, R.~Subramanian, and R.~Sundaresan, ``Interference planning for
  multicell {OFDM} downlink,'' in {\em IEEE Comsnets (invited)}, 2011.

\bibitem{veciana:LB}
H.~Kim, G.~de~Veciana, X.~Yang, and M.~Venkatachalam, ``Distributed
  $\alpha-$optimal user association and cell load balancing in wireless
  networks,'' {\em IEEE Trans. on Network.}, Feb. 2012.

\bibitem{madan:Rlb}
R.~Madan, J.~Borran, A.~Sampath, N.~Bhushan, A.~Khandekar, and T.~Ji, ``Cell
  association and interference coordination in heterogeneous {LTE-A} cellular
  networks,'' {\em IEEE J. Sel. Areas Comm.}, Dec. 2010.

\bibitem{yuw:UA}
K.~Shen and W.~Yu, ``Distributed pricing-based user association for downlink
  heterogeneous cellular networks,'' {\em IEEE Journal Sel. Areas. Commun.},
  Jun. 2014.

\bibitem{prasad:joint}
N.~Prasad, M.~Arslan, and S.~Rangarajan, ``Exploiting cell dormancy and load
  balancing in {LTE} hetnets: Optimizing the proportional fairness utility,''
  {\em IEEE Trans. on Commun.}, Oct. 2014.

\bibitem{ye:lb}
Q.~Ye, B.~Rong, Y.~Chen, M.~Al-Shalash, C.~Caramanis, and J.~Andrews, ``User
  association for load balancing in heterogeneous cellular networks,'' {\em
  IEEE Trans. on Wireless Comm.}, June 2013.

\bibitem{Buram:genpf}
T.~Bu, L.~Li, and R.~Ramjee, ``Generalized proportional fair scheduling in
  third generation wireless data networks,'' {\em IEEE Infocom}, 2006.

\bibitem{li:mrpf}
L.~Li, M.~Pal, and Y.~R. Yang, ``Proportional fairness in multi-rate wireless
  {LAN}s,'' {\em IEEE Infocom}, 2008.

\bibitem{Son:LB}
K.~Son, S.~Chong, and G.~Veciana, ``Dynamic association for load balancing and
  interference avoidance in multi-cell networks,'' {\em IEEE Trans. Wireless
  Comm.}, 2009.

\bibitem{bethan:arxiv}
D.~Bethanabhotla, O.~Bursalioglu, H.~Papadopoulos, and G.~Caire, ``Optimal
  user-cell association for massive mimo wireless networks,'' {\em v2, arXiv},
  feb 2015.

\bibitem{ehsan:rat}
E.~Aryafar, A.~Keshavarz-Haddad, M.~Wang, and M.~Chiang, ``{RAT} selection
  games in hetnets,'' in {\em IEEE Infocom}, 2013.

\bibitem{prasad:twoTS}
N.~Prasad, M.~Arslan, and S.~Rangarajan, ``A two time scale approach for
  coordinated multi-point transmission and reception over practical backhaul,''
  in {\em IEEE Comsnets (invited)}, jan 2014.

\bibitem{Yates:ULp}
R.~Yates, ``A framework for uplink power control in cellular radio systems,''
  {\em IEEE JSAC}, Sep. 1995.

\bibitem{Altman:Loc}
E.~Altman, A.~Kumar, C.~Singh, and R.~Sundaresan, ``Spatial sinr games of base
  station placement and mobile association,'' {\em IEEE Infocom.}, 2009.

\bibitem{bedekar:wiopt}
A.~Bedekar and R.~Agrawal, ``Optimal muting and load balancing for {eICIC},''
  in {\em {Proc. IEEE {WiOPT}}}, 2013.

\bibitem{hanly:eicic}
S.~Borst, S.~Hanly, and P.~Whiting, ``Throughput utility optimization in
  hetnets,'' in {\em IEEE VTC}, 2013.

\bibitem{siomina:lb}
I.~Siomina and D.~Yuan, ``Analysis of cell load coupling for {LTE} network
  planning and optimization,'' {\em IEEE Trans. on Wireless Comm.}, June 2012.

\bibitem{fehske:load}
A.~Fehske and G.~Fettweis, ``On flow level modeling of multi-cell wireless
  networks,'' in {\em IEEE WiOpt}, 2013.

\bibitem{linshroff:infocom}
X.~Lin and N.~Shroff, ``The impact of imperfect scheduling on cross-layer rate
  control in wireless networks,'' in {\em {Proc. IEEE {INFOCOM}}}, 2005.

\bibitem{deb:mota}
S.~Deb, A.~Keshavarz-Haddad, and V.~Srinivasan, ``{MOTA:} engineering an
  operator agnostic mobile service,'' in {\em IEEE Mobicom}, 2011.

\bibitem{rajiv_agarwal}
S.~Christensen, R.~Agarwal, E.~Carvalho, and J.~Cioffi, ``Weighted sum-rate
  maximization using weighted {MMSE} for {MIMO-BC} beamforming design,'' {\em
  IEEE Trans. Wireless Commun.}, vol.~7, no.~12, pp.~1--8, 2008.

\bibitem{stol:grad}
A.~L. Stolyar, ``On the asymptotic optimality of the gradient scheduling
  algorithm for multi-user throughput allocation,'' {\em Operations Res.},
  2005.

\bibitem{nemhaus:algo}
G.~L. Nemhauser and L.~A. Wolsey, ``Best algorithms for approximating the
  maximum of a submodular set function,'' {\em Math. Operations Research},
  1978.

\bibitem{lee:nmsub}
J.~Lee, V.~Mirrokni, V.~Nagarajan, and M.~Sviridenko, ``Non-monotone submodular
  maximization under matroid and knapsack constraints,'' in {\em STOC}, 2009.

\bibitem{goundan:sub}
P.~Goundan and A.~Schulz, ``Revisiting the greedy approach to submodular set
  function maximization,'' {\em manuscript}, June 2007.

\bibitem{zhang:dwr}
H.~Zhang, L.~Venturino, N.~Prasad, P.~Li, S.~Rangarajan, and X.~Wang,
  ``Weighted sum-rate maximization in multi-cell networks via coordinated
  scheduling and discrete power control,'' {\em IEEE J. Sel. Areas Comm.}, Dec.
  2010.

\bibitem{palomar:jsac}
D.~P. Palomar and M.~Chiang, ``A tutorial on decomposition methods for network
  utility maximization,'' {\em IEEE {JSAC}}, Aug. 2006.

\bibitem{book:chiangGP}
M.~Chiang, {\em Geometric Programming for Communication Systems}.
\newblock Now Publishers, 2005.

\end{thebibliography}
